\documentclass[a4paper,twocolumn,11pt,accepted=2021-07-26]{quantumarticle}
\pdfoutput=1
\usepackage[utf8]{inputenc}
\usepackage[english]{babel}
\usepackage[T1]{fontenc}
\usepackage{amsmath,amssymb}
\usepackage{url}
\usepackage{hyperref}
\usepackage[numbers,sort&compress]{natbib}

\usepackage{tikz}
\usepackage{lipsum}
\usepackage{braket}
\usepackage{listings}
\usepackage{subfigure}
\usepackage{bm}

\definecolor{codegreen}{rgb}{0,0.6,0}
\definecolor{codegray}{rgb}{0.5,0.5,0.5}
\definecolor{codepurple}{rgb}{0.58,0,0.82}
\definecolor{backcolour}{rgb}{0.95,0.95,0.92}

\lstdefinestyle{mystyle}{
	backgroundcolor=\color{backcolour},   
	commentstyle=\color{codegreen},
	keywordstyle=\color{magenta},
	numberstyle=\tiny\color{codegray},
	stringstyle=\color{codepurple},
	basicstyle=\ttfamily\footnotesize,
	breakatwhitespace=false,         
	breaklines=true,                 
	captionpos=b,                    
	keepspaces=true,                 
	numbers=left,                    
	numbersep=5pt,                  
	showspaces=false,                
	showstringspaces=false,
	showtabs=false,                  
	tabsize=2
}
\begin{document}
\title{Qulacs: a fast and versatile quantum circuit simulator for research purpose}

\author{Yasunari Suzuki}
\email{yasunari.suzuki.gz@hco.ntt.co.jp}
\affiliation{NTT Computer and Data Science Laboratories, NTT Corporation, Musashino 180-8585, Japan}
\affiliation{JST PRESTO, Kawaguchi, Saitama 332-0012, Japan}

\author{Yoshiaki Kawase}
\affiliation{Graduate School of Engineering Science, Osaka University, 1-3 Machikaneyama, Toyonaka, Osaka 560-8531, Japan}

\author{Yuya Masumura}
\affiliation{Graduate School of Information Science and Technology, Osaka University, 1-1 Yamadaoka, Suita, Osaka 565-0871, Japan}
\author{Yuria Hiraga}
\affiliation{Graduate School of Information and Science, Nara Institute of Science and Technology, Takayama, Ikoma, Nara 630-0192, Japan}
\author{Masahiro Nakadai}
\affiliation{Graduate School of Science, Kyoto University, Yoshida-Ushinomiya, Sakyo, Kyoto 606-8302, Japan}

\author{Jiabao Chen}
\email{qulacs@qunasys.com}
\affiliation{QunaSys Inc., Aqua Hakusan Building 9F, 1-13-7 Hakusan, Bunkyo, Tokyo 113-0001, Japan}
\author{Ken M. Nakanishi}
\affiliation{QunaSys Inc., Aqua Hakusan Building 9F, 1-13-7 Hakusan, Bunkyo, Tokyo 113-0001, Japan}
\affiliation{Graduate School of Science, The University of Tokyo, 7-3-1 Hongo, Bunkyo-ku, Tokyo 113-0033, Japan}
\author{Kosuke Mitarai}
\affiliation{Graduate School of Engineering Science, Osaka University, 1-3 Machikaneyama, Toyonaka, Osaka 560-8531, Japan}
\affiliation{QunaSys Inc., Aqua Hakusan Building 9F, 1-13-7 Hakusan, Bunkyo, Tokyo 113-0001, Japan}
\affiliation{Center for Quantum Information and Quantum Biology, Institute for Open and Transdisciplinary Research Initiatives, Osaka University, Japan}
\author{Ryosuke Imai}
\email{qulacs@qunasys.com}
\affiliation{QunaSys Inc., Aqua Hakusan Building 9F, 1-13-7 Hakusan, Bunkyo, Tokyo 113-0001, Japan}
\author{Shiro Tamiya}
\affiliation{QunaSys Inc., Aqua Hakusan Building 9F, 1-13-7 Hakusan, Bunkyo, Tokyo 113-0001, Japan}
\affiliation{Graduate School of Engineering, The University of Tokyo, 7-3-1 Hongo, Bunkyo-ku, Tokyo 113-0033, Japan}
\author{Takahiro Yamamoto}
\affiliation{QunaSys Inc., Aqua Hakusan Building 9F, 1-13-7 Hakusan, Bunkyo, Tokyo 113-0001, Japan}
\author{Tennin Yan}
\email{qulacs@qunasys.com}
\affiliation{QunaSys Inc., Aqua Hakusan Building 9F, 1-13-7 Hakusan, Bunkyo, Tokyo 113-0001, Japan}
\author{Toru Kawakubo}
\email{qulacs@qunasys.com}
\affiliation{QunaSys Inc., Aqua Hakusan Building 9F, 1-13-7 Hakusan, Bunkyo, Tokyo 113-0001, Japan}
\author{Yuya O. Nakagawa}
\email{qulacs@qunasys.com}
\affiliation{QunaSys Inc., Aqua Hakusan Building 9F, 1-13-7 Hakusan, Bunkyo, Tokyo 113-0001, Japan}
\author{Yohei Ibe}
\affiliation{QunaSys Inc., Aqua Hakusan Building 9F, 1-13-7 Hakusan, Bunkyo, Tokyo 113-0001, Japan}
\author{Youyuan Zhang}
\affiliation{QunaSys Inc., Aqua Hakusan Building 9F, 1-13-7 Hakusan, Bunkyo, Tokyo 113-0001, Japan}
\affiliation{Graduate School of Science, The University of Tokyo, 7-3-1 Hongo, Bunkyo-ku, Tokyo 113-0033, Japan}

\author{Hirotsugu Yamashita}
\affiliation{Individual researcher}
\author{Hikaru Yoshimura}
\affiliation{Individual researcher}

\author{Akihiro Hayashi}
\affiliation{School of Computer Science, Georgia Institute of Technology, Atlanta, GA, 30332, USA}
\author{Keisuke Fujii}
\affiliation{JST PRESTO, Kawaguchi, Saitama 332-0012, Japan}
\affiliation{Graduate School of Engineering Science, Osaka University, 1-3 Machikaneyama, Toyonaka, Osaka 560-8531, Japan}
\affiliation{Center for Quantum Information and Quantum Biology, Institute for Open and Transdisciplinary Research Initiatives, Osaka University, Japan}
\affiliation{Center for Emergent Matter Science, RIKEN, Wako Saitama 351-0198, Japan}

\begin{abstract}
To explore the possibilities of a near-term intermediate-scale quantum algorithm and long-term fault-tolerant quantum computing, a fast and versatile quantum circuit simulator is needed. Here, we introduce Qulacs, a fast simulator for quantum circuits intended for research purpose. We show the main concepts of Qulacs, explain how to use its features via examples, describe numerical techniques to speed-up simulation, and demonstrate its performance with numerical benchmarks.
\end{abstract}

\maketitle

\section{Introduction}
Many theoretical groups have explored quantum computing applications due to the rapid improvements in quantum technologies and huge efforts of experimental groups to develop quantum computers~\cite{arute2019quantum,egan2020fault}. Although classical simulation of quantum circuits is a vital tool to develop quantum computers, the simulation time increases exponentially with the number of qubits. 

The primary reasons for using a classical simulator are the following:
(i) Quantum devices suffer from much higher error rates. A classical simulator is necessary to determine the ideal results for comparison. 
(ii) In certain cases, not all necessary values can be directly measured from experiments such as the full quantum state vector and marginal probabilities of measurements. 
(iii) To analyze the performance of quantum error correction and noise characterization for an arbitrary noise model, noisy quantum circuits must be simulated. 
Hence, there are broad demands on classical simulators for research on quantum computing. 
Although full simulations of quantum circuits are not efficient in the sense of computational complexity theory, it is important to implement a fast classical simulator as much as possible.

Here, we introduce a quantum circuit simulator, which is called {\it Qulacs}~\cite{Qulacs}. 
The main feature of Qulacs is that it meets many popular demands in quantum computing research such as evaluating near-term applications, quantum error correction, and quantum benchmark methods. In addition, Qulacs is available on many popular environments, and is one of the fastest quantum circuit simulators. Herein we demonstrate the structure of our library, optimization methods, and numerical benchmarks for simulating typical quantum gates and circuits.

This paper consists of the following topics. Section.\,\ref{sec:overview} overviews the main features and structure of Qulacs as well as compares it with existing libraries. Section.\,\ref{sec:expected_use} introduces the expected use of Qulacs. Section.\,\ref{sec:implementation} discusses how to write codes with Qulacs using examples codes. Section.\,\ref{sec:optimization} introduces further optimization techniques to speed-up the simulation of quantum circuits. Section.\,\ref{sec:performance} shows the numerical benchmarks for Qulacs. Section\,\ref{sec:comparison} compares the performance of Qulacs with that of existing simulators. Finally, Section.\,\ref{sec:conclusion} is devoted to summary and discussion.

\section{Overview}
\label{sec:overview}
\subsection{Features of Qulacs}
Qulacs is designed to accelerate research on quantum computing. Thus, Qulacs prioritizes the following: 

\paragraph{Fast simulation of large quantum circuits:}
A full simulation of quantum circuits requires a time that grows exponentially with the number of qubits. This problem can be mitigated by optimizing and parallelizing the simulation codes for single- or multi-core CPUs and SIMD (Single Instruction Multiple Data) units and even GPUs (Graphics Processing Units), and optimizing quantum circuits before a simulation. These techniques can enable a few orders of magnitude performance improvement compared to naive implementations. Although this is a constant factor speed-up, the effect on practical research is tremendous. Qulacs offers optimized and parallelized codes to update a quantum state and evaluate probability distributions, observables, and so on.

\paragraph{Small overhead for simulating small quantum circuits:}
Often small but noisy quantum circuits (up to 10 qubits) are simulated many times rather than simulating a single-shot large ideal quantum circuit. 
In this case, the overhead due to pre- and post-processing for calling core API functions is not negligible relative to the overall simulation time. Qulacs is designed to minimize such overhead by focusing on core features and avoiding complicated functionalities. 

\paragraph{Available on different environments:}
While numerical analysis is typically performed on workstations or high-performance computers, most software development occurs on laptops or desktop personal computers. For versatility, Qulacs provides interfaces for both Python and C++ languages, while most of the codes of Qulacs are written in C and C++ languages. In addition, Qulacs supports several compilers such as GNU Compiler Collection (GCC) and Microsoft Visual Studio C++ (MSVC). Qulacs is tested on several operating systems such as Linux, Windows, and Mac OS. 

\paragraph{Many useful utilities for research:}
To meet various demands in quantum computing research, a simulator should support general quantum operations. Qulacs can create not only common unitary gates and projection measurements, but also general operations such as completely positive instruments and adaptive quantum gates conditioned on measurement results. 

\subsection{Structure of Qulacs}
Qulacs consists of three shared libraries. The first one is a core library written in C language for optimized memory management, update functions of quantum states, and property evaluation of quantum states. The second library is built on top of the first library, and is written in C++ language. This allows users to easily create and control quantum gates and circuits in an object-oriented way, thereby improving the programmability. Also, at runtime, it adaptively chooses the best performing implementation of quantum gates depending on the number of qubits. The third library explores variational methods with quantum circuits. This library wraps quantum gates and circuits so that quantum circuits can be treated as variational objects. 

\begin{figure*}[t]
	\centering
    \includegraphics[width=\textwidth, clip]{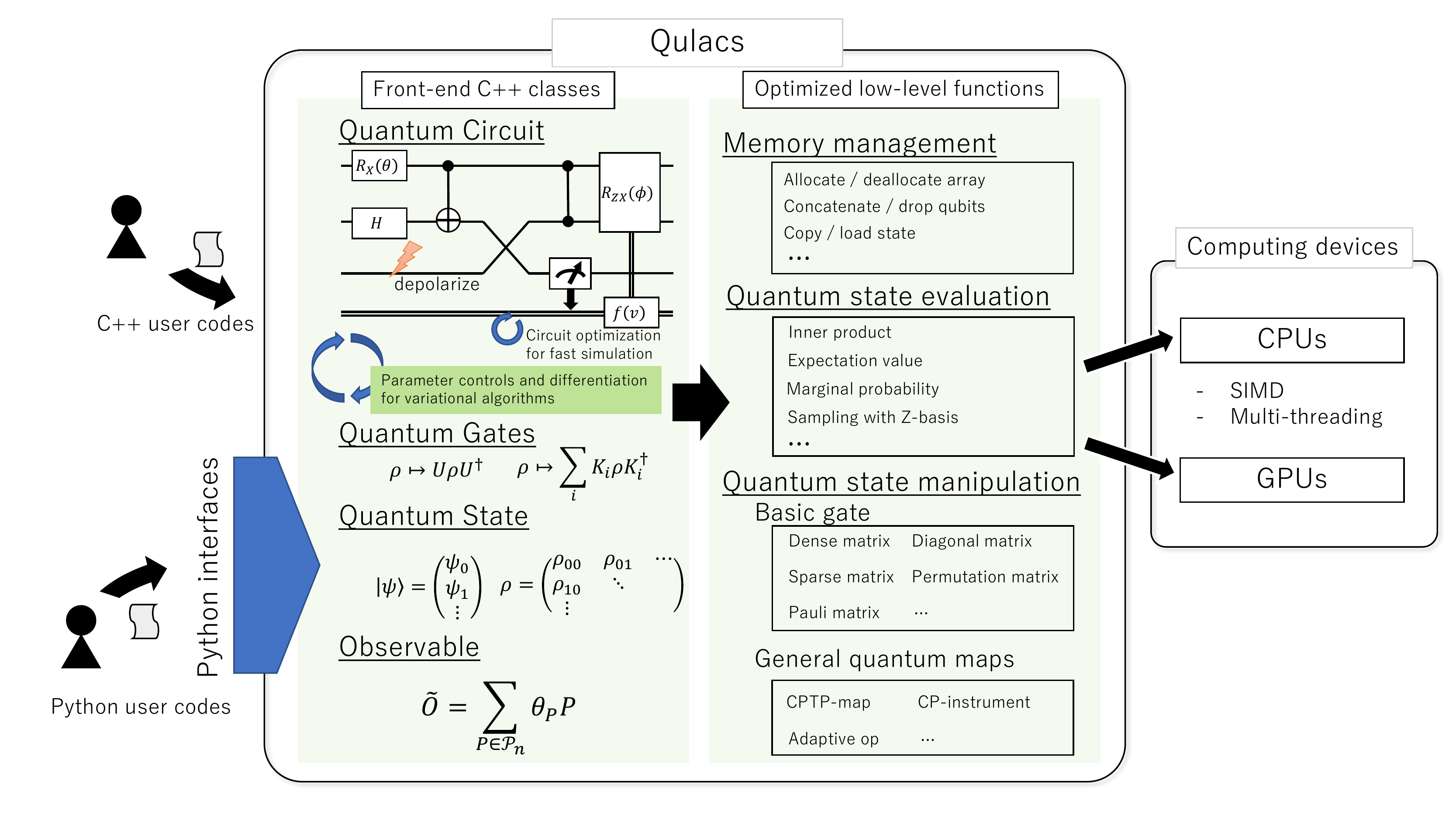}
	\caption{The overview of the structure of Qulacs.}
    \label{fig:overview}
\end{figure*}

We expect that users who work on variational quantum algorithms use Qulacs via the third library, and the other users access Qulacs via the second library. Additionally, Qulacs can be used as a Python library, while there is some overhead in interfacing between Python and C++. Figure\,\ref{fig:overview} shows the overview of the structure of Qulacs. The components of Qulacs and their usages are explained in Sec.\,\ref{sec:implementation} with example codes.
Qulacs uses Eigen~\cite{eigenweb} to treat a sparse matrix and to manage matrix representations of quantum gates and pybind11~\cite{pybind11} for exporting functions and classes from C++ to python. The features introduced in this paper is fully tested with GoogleTest~\cite{googletest} and pytest~\cite{pytestx.y}.

\subsection{Simulation methods}
There are several approaches to simulate general quantum circuits with classical computers. The simplest approach is to update quantum states represented by state vectors or density matrices sequentially by applying quantum gates as general maps. This method is called Schr\"{o}dinger's method~\cite{arute2019quantum}. 
Qulacs implements this method for simulating quantum circuits due to its fast and versatile simulations of quantum circuits compared with the other methods introduced in this section. A detail of implementation with this method is described in Sec.\,\ref{sec:implementation}. 

Another method is Feynman's approach~\cite{arute2019quantum,boixo2017simulation,markov2008simulating}, which computes the sum of all Feynman's path contributions. This technique greatly decreases the memory size requirement, allowing for the single amplitude of the final quantum state to be quickly known. 
While the number of Feynman's paths increases exponentially according to not only the number of qubits but also the number of quantum gates, this requirement can be relaxed by using tensor-network-based approach. With tensor-network-based simulator, we can make the simulation time increases exponentially according to a tree-width of the network~\cite{markov2008simulating}, which is a characteristic value of graph representing how a tensor network is close to a tree graph. 
However, except special cases where tree-width is small, such as shallow quantum circuits with a large number of qubits, a tree-width becomes equal to the number of qubits. Moreover, the Schr\"{o}dinger's method is much faster than Feynman's path integral when the number of qubits is limited and the memory size is sufficient for storing a full state vector. 
Although we can also consider Schr\"{o}dinger-Feynman approach~\cite{markov2018quantum,arute2019quantum} as a hybrid method, this method is typically faster than Schr\"{o}dinger's method simulating large quantum circuits with a small depth.

If quantum circuits have specific features, then the simulation time can be reduced. For example, if quantum circuits are dominated by Clifford gates, non-Clifford gates can be treated as a perturbation to the simulation. This treatment can reduce the time for simulation~\cite{Bravyi2016Improved,Bravyi2019simulationofquantum}. This condition is satisfied in several situations, e.g., quantum circuits of stabilizer measurements where non-Clifford errors happens with a very small probability or fault-tolerant quantum computing with a limited number of $T$ and TOFFOLI-gates. However, most of the quantum circuits of typical quantum algorithms do not satisfy these conditions. 

\subsection{Relation to the existing libraries}
To date, many groups have published a variety of quantum circuit simulators. Cirq~\cite{quantum_ai_team_and_collaborators_2020_4062499}, Qiskit~\cite{Qiskit}, PyQuil~\cite{smith2016practical}, and PennyLane~\cite{bergholm2018pennylane} are published by Google, IBM, Rigetti computing, and Xanadu, respectively. Since these groups are developers of hardware for quantum computing, these libraries are designed for submitting quantum tasks as a job without thinking about detailed experimental procedures or setups. Q\#~\cite{10.1145/3183895.3183901} by Microsoft focuses on providing higher levels of abstraction, such as packaged instructions for integer arithmetic or tool-chains for compilation. To simulate quantum circuits with a low depth and a large number of qubits, tensor-network-based simulators are used~\cite{Villalonga2019,roberts2019tensornetwork,itensor}. However, these are not good for higher depths or a fewer number of qubits. For quantum circuit simulations on state-of-the-art supercomputers, several works have reported on the performance and optimizations~\cite{villalonga2020establishing,de2007massively,de2019massively,haner20175}.
Qiskit-Aer~\cite{Qiskit}, Intel-QS~\cite{guerreschi2020intel,smelyanskiy2016qhipster}, QX~Simulator~\cite{khammassi2017qx,khammassi2020openql}, ProjectQ~\cite{steiger2018projectq}, QuEST~\cite{jones2019quest}, qsim~\cite{quantum_ai_team_and_collaborators_2020_4023103}, Yao~\cite{Luo2020yaojlextensible}, QCGPU~\cite{kelly2018simulating}, and Qibo~\cite{efthymiou2020qibo} were all developed with motivations similar to ours. They focused on optimizing quantum circuit simulations for quantum computing researchers. By contrast, Qulacs is one of the fastest simulators, which minimizes overhead even for small simulations, supports general quantum operations such as completely-positive and trace-preserving maps and completely-positive instruments, and runs on various environments. In addition, Qulacs provides many useful utilities frequently used in research such as calculations of transition amplitudes and reversible Boolean functions.

\section{Expected usages of Qulacs}
\label{sec:expected_use}
Quantum circuit simulators should be designed for specific targets. Qulacs is designed to help researchers of quantum computing. In particular, we expect the following usages: 

\subsection{Exploration of near-term applications and error mitigation techniques}
To highlight typical evaluation targets, here we show two popular directions of near-term applications. One is optimizing a target function using variational quantum circuits such as the variational quantum eigensolver (VQE) ~\cite{peruzzo2014variational}. In a typical scenario, we assume rotation angles of Pauli gates as variational parameters of a cost function and optimize them by repeatedly simulating quantum circuits with a relatively small number of qubits. The other application is a quantum simulator~\cite{lloyd1996universal}. In this approach, large quantum systems are simulated to explore physics in many-body quantum systems. In a typical scenario, we perform a single simulation with quantum circuits as large as possible. Thus, the size of memory or allowed time limits the size of quantum circuits. 

To date, Qulacs has been used in a few tens of research papers. For example, Qulacs is used by papers related to noisy intermediate-scale quantum (NISQ) applications~\cite{endo2020calculation, endo2020calculation, mitarai2020theory,mitarai2019generalization,matsuzawa2020jastrow,Kawai_2020,kottmann2021quantum} and fault-tolerant quantum computing~\cite{suzuki2020quantum}.
Although Qulacs does not support gate decomposition, which is supported by high-layer libraries, Qulacs can be used as a faster backend library. For example, Qulacs can serve as a backend of Cirq~\cite{quantum_ai_team_and_collaborators_2020_4062499} using a library cirq-qulacs~\cite{CirqQulacs}. Qulacs has also been chosen as a fast backend simulator in several libraries and services such as PennyLane (Xanadu)~\cite{bergholm2018pennylane}, t$|$ket$>$ (Cambridge Quantum Computing)~\cite{sivarajah2020t}, Orquestra (Zapata computing) ~\cite{Orquestra}, and Tequila~\cite{tequila}.

\subsection{Performance analysis of quantum error correction schemes}
Another important usage of the simulator for quantum circuits is performance analysis of the quantum error correction and fault-tolerant quantum computing. To construct a quantum computer large enough for Shor’s algorithm ~\cite{shor1999polynomial,gidney2021factor}, a quantum simulation for quantum many-body systems~\cite{lloyd1996universal,kivlichan2020improved}, or algorithms for linear systems~\cite{harrow2009quantum}, quantum error correction~\cite{fowler2012surface} is necessary to reduce logical error rates to an arbitrarily small value. 
Many types of quantum error-correcting codes and schemes have been proposed. However, the number of qubits needed for a specific application is highly dependent on the performance of quantum error-correcting schemes. Consequently, it is difficult to control the noise properties on real devices and the performance of quantum error correction must be analyzed with classical simulation in near-term development. Unfortunately, the time to accurately simulate quantum error-correcting codes with practical noise models grows exponentially with the number of qubits. Thus, we need a fast and accurate simulator of noisy quantum circuits of quantum error correction. 

\subsection{Generation of a reference of experimental data}
To characterize and calibrate controls of qubits, sometimes experimental data must be compared with the ideal one. For example, several verification methods for large quantum devices~\cite{boixo2018characterizing,arute2019quantum} require a full simulation of large quantum systems. While quantum circuits in quantum computational supremacy regime require supercomputers for simulations, portable and fast quantum circuit simulators remain useful for generating small-scale experimental references.

\section{Implementation of Qulacs}
\label{sec:implementation}
\subsection{Overview}
In Qulacs, any state of a quantum system is represented as a subclass of the \texttt{QuantumStateBase} class. Thus far, Qulacs supports two representations of quantum states: state vector and density matrix. The \texttt{StateVector} class represents a state vector, while the \texttt{DensityMatrix} class represents a density matrix. 
These classes have some basic utilities as their member functions such as initialization to a certain quantum state, computing marginal probabilities, and sampling measurement results.
The \texttt{QuantumStateBase} class also contains a variable-length integer array called classical registers, which are used to store measurement results. 

When a quantum state is updated by quantum operations, subclasses of \texttt{QuantumGateBase} are instantiated and applied to a quantum state. This class supports not only unitary operations and projection measurements but also a variety of operations for general quantum mapping such as a completely-positive instrument and a completely-positive trace-preserving map. 

To evaluate the expectation values of observables, there is a class named \texttt{Observable}. We assume the \texttt{Observable} class is described as a linear combination of Pauli operators. Thus, the \texttt{Observable} instance can be constructed directly or from an output of an external library such as OpenFermion~\cite{mcclean2020openfermion}. Additionally, a Trotterized quantum circuit can be created from a given observable.

By default, Qulacs performs a simulation by allocating and manipulating \texttt{StateVector} on a CPU. Also, since the use of GPUs can significantly outperform a CPU in certain cases, Qulacs supports GPU execution. This can be done by using the \texttt{StateVectorGpu}. 
Once a state vector is allocated on a GPU, all the computations such as update and evaluation of state vectors are performed in a GPU unless the state is explicitly converted into \texttt{StateVector}. Qulacs does not support allocating a quantum state on multiple GPUs. One GPU can be selected by supplying an index if multiple GPUs are installed.

In the discussion below, we explain the case where quantum states are allocated as state vectors on the main RAM and processed within a CPU. Although Qulacs can be used as a C++ library, we show examples in the Python language for simplicity in the main text. See Appendix.\,\ref{appendix:impl_cpp} for the example codes of the C++ language.
Here, we show typical examples and their basic features. Qulacs supports more operations than those explained here. For a detailed explanation, please see the documentation on the official website~\cite{Qulacs}. 

\subsection{Quantum state}
\subsubsection{Initialization}
In Qulacs, instantiating the \texttt{StateVector} class allocates a state vector. By default, a state vector is initialized to zero-state $\ket{0}^{\otimes n}$. However, the state can be initialized to other states such as a computational basis, a state vector of a given complex array, or a random pure state with its member functions. The instance of the \texttt{StateVector} class can create its copy or load the contents of another \texttt{StateVector}. Listing.\,\ref{alg_init} shows example codes.

\begin{lstlisting}[language=Python, caption=An example Python program that initializes quantum states., label=alg_init]
import numpy as np
from qulacs import StateVector

# Allocate a state vector
num_qubit = 2
state = StateVector(num_qubit)

# Reset to a computational basis
## (0:|00>, 1:|01>, 2:|10>, 3:|11>)
## Note that the right-most digit corresponds to
##  the 0-th qubit in Qulacs.
state.set_computational_basis(index = 2)

# Create a copy of the state vector
sub_state = state.copy()

# Load a given list, numpy array, 
#  or another StateVector
state.load(state=[0.5, 0.5, 0.5, -0.5])
state.load(np.ones(4)/2)
state.load(sub_state)

# Prepare a randomized pure quantum state
state.set_Haar_random_state(seed = 42)
\end{lstlisting}

\subsubsection{Analysis}
Qulacs implements several functions to evaluate the properties of quantum states.
Although the \texttt{get\_vector} functions provide a full state vector, evaluating the properties with built-in functions is fast. For example, Listing.\,\ref{alg_char} shows example codes to compute a marginal probability, squared norm, and inner-product of two states. Note that the \texttt{sampling} functions create a cumulative probability distribution as pre-processing for the fast sampling, which temporally allocates an additional $2^n$-length array. 

\begin{lstlisting}[language=Python, caption=An example Python program that evaluates the properties of quantum states., label=alg_char]
from qulacs import StateVector
num_qubit = 3
state = StateVector(num_qubit)
state.set_Haar_random_state(0)

# Get the state vector as numpy array
vec = state.get_vector()

# Get the marginal probability
## The below example obtains prob of |021>
## where "2" is a wild card 
## matching to both 0 and 1.
prob = state.get_marginal_probability(measured_value=[1, 2, 0])

# Sampling results of the Z-basis measurements
samples = state.sampling(count=100, seed=42)

# Computing the squared norm
squared_norm = state.get_squared_norm()

# Computing the inner product of two quantum states
from qulacs.state import inner_product
state_bra = StateVector(num_qubit)
state_bra.set_Haar_random_state()
state_ket = StateVector(num_qubit)
state_ket.set_Haar_random_state()
value = inner_product(state_bra, state_ket)
\end{lstlisting}

\subsubsection{Update}
Several member functions of \texttt{StateVector} can be used for quickly updating a state vector. 
The \texttt{multiply\_coef} function multiplies a complex number to each element of a quantum state, while the \texttt{multiply\_elementwise\_function} multiplies index-dependent coefficients to a state vector with a function that returns a coefficient according to a given index. 
The \texttt{add\_state} function adds two state vectors. While these operations are not physically achievable, they are useful for analysis in theoretical studies such as creating a superposition of two given states and supplying specific phases to each element of a state vector. A list of qubits in a state vector can be concatenated, permutated, or reduced with \texttt{tensor\_product}, \texttt{permutate\_qubit}, or \texttt{drop\_qubit}, respectively. Listing.\,\ref{alg_mod} shows code examples for multiplication kernels.

\begin{lstlisting}[language=Python, caption=An example Python program that update and modify quantum states., label=alg_mod]
from qulacs import StateVector
from qulacs.state import tensor_product, permutate_qubit, drop_qubit

state = StateVector(2)
state.set_Haar_random_state()

# Normalize the state vector
squared_norm = state.get_squared_norm()
state.normalize(squared_norm)

# Multiply a complex number to each element
state.multiply_coef(0.5+0.1j)

# Perform element-wise multiply
def func(index):
  return (0.5 if index%2 else 0)

state.multiply_elementwise_function(func)

# Perform element-wise addition
state.add_state(state)

# Make a tensor product of states.
#  The resultant state has 4 qubits
sub_state = StateVector(2)
state = tensor_product(state, sub_state)

# Permutate qubit indices from [0,1,2,3] to [3,1,2,0]
state = permutate_qubit(state, [3,1,2,0])

# Drop the 1-st and 2-nd qubits from state
#  and project to |0> and |0> subspace.
new_state = drop_qubit(state, [1,2], [0,0])

\end{lstlisting}

\subsection{Quantum gates}
\subsubsection{Gate type}
As explained in the overview, quantum gates include not only typical quantum gates such as unitary operators and Pauli-$Z$ basis measurements, but also contain all operations that update quantum states. All the classes of quantum gates are defined as a subclass of the \texttt{QuantumGateBase} class, which has a function \texttt{update\_quantum\_state} that acts on derived classes of \texttt{QuantumStateBase}.
In Qulacs, quantum gates in which the action can be written as $\ket{\psi} \mapsto K \ket{\psi}$, where $\ket{\psi}$ is a state vector and $K$ is a certain complex matrix, are called basic gates. Examples include a Pauli rotation on multiple qubits, TOFFOLI-gate, and stabilizer projection to $+1$ eigenspace. 
Quantum maps that are not basic gates such as CPTP-map, projection measurements, and adaptive operations, are represented using basic gates. 
Here, we show several popular types of operations which are implemented as basic operations.

\subsubsection{Basic gate}
A basic gate is an operation that can be represented by $\rho \mapsto K \rho K^{\dagger}$. In the case of a pure state, its action on the state vector is represented by $\ket{\psi} \mapsto K \ket{\psi}$. Note that $K$ is not necessarily unitary. 
Suppose that this quantum gate acts on a pure quantum state $\ket{\psi}$ and obtain an updated quantum state $\ket{\psi'}=K\ket{\psi}$. We denote a state vector of the computational basis as $\ket{\bm{x}} = \bigotimes_i \ket{x_i}$ for $\bm{x} \in \{0,1\}^n$, a coefficient of the state vector along with the computational basis as $\psi_{\bm{x}}=\braket{\bm{x} | \psi}$, and a matrix representation of $K$ as $K_{\bm{x}, \bm{y}} = \braket{\bm{x}|K|\bm{y}}$.
Then, a coefficient of the updated quantum state can be written by
\begin{align}
\label{eq:update_raw}
\psi'_{\bm{x}} = \sum_{\bm{y} \in \{0,1\}^n} K_{\bm{x}, \bm{y}} \psi_{\bm{y}}.
\end{align}

Typically, quantum gates non-trivially act only on a few qubits. Suppose that the total number of qubits is $n$, the list of qubits on which the quantum gate non-trivially acts is $M$, and the number of the target qubits is $m$.
Then, the action of the quantum gate $K$ can be simplified by using a $2^m \times 2^m$ complex matrix $\tilde{K}$, which we call a gate matrix, as follows.
We define the following two lists of $n$-bit strings:
\begin{align}
B^{(0)} = \{ \bm{x} \in \{0,1\}^n | \forall i \in M, x_i = 0\} \\
B^{(1)} = \{ \bm{x} \in \{0,1\}^n | \forall i \notin M, x_i = 0\}.
\end{align}
Here, $B^{(0)}$ is a list of $n$-bit strings where all the values at indices contained in $M$ are zero, and $B^{(1)}$ is a list of $n$-bit strings where all the values at indices not contained in $M$ are zero. There are $2^{n-m}$ elements in $B^{(0)}$ and $2^m$ elements in $B^{(1)}$. An arbitrary $n$-bit string $\bm{x}$ can be uniquely decomposed as $\bm{x} = \bm{x}^{(0)} + \bm{x}^{(1)}$ where $\bm{x}^{(i)} \in B_i$. We use this decomposition implicitly in the following discussion. Since the quantum gate only acts on the target qubits, a transition amplitude between two computational basis states is zero if their $n$-bit strings are different at indices not contained in the target qubits, and a transition amplitude is independent of the values at indices not contained in the target qubit, i.e.,
\begin{align}
K_{\bm{x}, \bm{y}} &= K_{\bm{x}^{(0)}+\bm{x}^{(1)}, \bm{y}^{(0)}+\bm{y}^{(1)}} \nonumber \\
&= K_{\bm{x}^{(1)}, \bm{y}^{(1)}} \delta_{\bm{x}^{(0)}, \bm{y}^{(0)}}
\end{align}
for an arbitrary $\bm{x}, \bm{y} \in \{0,1\}^n$. 
Then, Eq.\,(\ref{eq:update_raw}) can be rephrased as 
\begin{align}
\label{eq:update_decomp}
\psi'_{\bm{x}^{(0)} + \bm{x}^{(1)}} = \sum_{\bm{y}^{(1)} \in B^{(1)}} K_{\bm{x}^{(1)}, \bm{y}^{(1)}} \psi_{\bm{x}^{(0)} + \bm{y}^{(1)}}.
\end{align}
We define a bijective function $\bm{r}: B^{(1)} \rightarrow \{0,1\}^m$ such that $(x_0, \cdots, x_{n-1}) \mapsto (x_{M_0}, \cdots, x_{M_{m-1}})$ where $M_i$ is the qubit index of the $i$-th target qubit, and also define $\bm{s}: \{0,1\}^m \rightarrow B^{(1)}$ as the inverse of $\bm{r}$. Then, a $2^m \times 2^m$ gate matrix $\tilde{K}$ of the quantum gate $K$ is defined as
\begin{align}
\tilde{K}_{\bm{z}, \bm{w}} = \braket{\bm{s}(\bm{z}) | K | \bm{s}(\bm{w})}
\end{align}
for $\bm{z}, \bm{w} \in \{0,1\}^m$. With the gate matrix, we can write Eq.\,(\ref{eq:update_decomp}) as follows:
\begin{align}
\label{eq:update}
\psi'_{\bm{x}^{(0)} + \bm{s}(\bm{z})} = \sum_{\bm{w} \in \{0,1\}^m} \tilde{K}_{\bm{z}, \bm{w}} \psi_{\bm{x}^{(0)} + \bm{s}(\bm{w})}.
\end{align}
We also define a temporal state vector $\ket{\tilde{\psi}_{\bm{x}^{(0)}}}$ as a $2^m$-dim complex vector of which the $i$-th element is $\psi_{\bm{x}^{(0)} + \bm{s}({\rm bin}(i))}$ where ${\rm bin}(i)$ represents the $m$-bit binary representation of the integer $i$. Then, we can simplify Eq.\,(\ref{eq:update}) as \begin{align}
\label{eq:update_mat}
\ket{\tilde{\psi'}_{\bm{x}^{(0)}}} = \tilde{K} \ket{\tilde{\psi}_{\bm{x}^{(0)}}}.
\end{align}
Since for all $\bm{x}^{(0)} \in B^{(0)}$, Eq.\,(\ref{eq:update_mat}) has to be calculated, an update function for $K$ consists of $2^{n-m}$ matrix-vector multiplications with a gate matrix $\tilde{K}$ and the temporal state vectors $\ket{\tilde{\psi}_{\bm{x}^{(0)}}}$. 
Listing.\,\ref{alg_pseudo} shows a naive implementation of update functions in the Python language, where \texttt{B0} and \texttt{B1} are a list of $n$-bit integers corresponding to $B_0$ and $B_1$, respectively. 
\begin{lstlisting}[language=Python, caption=An example Python code that naively implements an update function of a quantum state, label=alg_pseudo]
import numpy as np

def func(state_vector, gate_matrix, B0, B1, m):
  temp_vector = np.zeros(2**m)
  for x0 in B0:
    # Read values from the state vector
    for ind, x1 in enumerate(B1):
      temp_vector[ind] = state_vector[x0+x1]
    # Perform matrix-vector multiplication
    temp_vector = np.dot(gate_matrix, temp_vector)
    # Write values to the state vector
    for ind, x1 in enumerate(B1):
      state_vector[x0+x1] = temp_vector[ind]
\end{lstlisting}
Note that in the example code, without loss of generality, $B_1$ is considered to be arranged so that the $i$-th element of $B_1$ is $\bm{s}({\rm bin}(i))$. 
In practice, a time for reading/writing temporal state vectors (i.e., \texttt{temp\_vector} in the example code) from/to the whole state vector is not negligible. Thus, the update function essentially performs $2^{n-m}$ iterations of the following three parts: read $2^m$-dim complex numbers from a memory, perform matrix-vector multiplication, and write $2^m$ complex numbers to the memory.

The \texttt{DenseMatrix} function generates a basic gate $K$ with a gate matrix $\tilde{K}$, and the \texttt{RandomUnitary} function generates that with a random unitary matrix sampled from a Haar-random distribution. Listing.\,\ref{alg_dense} shows quantum gates instantiated with dense matrices.

\begin{lstlisting}[language=Python, caption=An example Python program that applies dense matrix gates to quantum states., label=alg_dense]
from qulacs import StateVector
from qulacs.gate import DenseMatrix, RandomUnitary
state = StateVector(10)

# Update quantum state with given gate matrix
target_list = [1]
gate_matrix = [[0, 1],[1, 0]]
gate = DenseMatrix(target_list, gate_matrix)
gate.update_quantum_state(state)

# Update quantum state with random unitary
## Matrix is drawn from Haar-random distribution
random_gate = RandomUnitary(target_list)
random_gate.update_quantum_state(state)
\end{lstlisting}

The computation time to apply a quantum gate is mainly determined by two factors: time for processing arithmetic operations to perform a calculation with complex numbers and time for processing memory operations to transfer complex numbers between the CPU and main RAM; we call these arithmetic-operation cost and memory-operation cost, respectively. They are proportional to the number of arithmetic and memory operations in update functions.
The heavier of the two determines the application time of a function. 
The number of arithmetic operations in each iteration of an update function of dense matrix gates is $O(2^{2m})$ and that of memory operations is $O(2^{m} + 2^{2m})$, where $O$ is a Landau notation. Since this iteration is looped for $2^{n-m}$ times, the total number of arithmetic and memory operations are $O(2^{n+m})$ and $O(2^{n} + 2^{n+m})$. In the case of small $m$, the gate matrix $\tilde{K}$, which is re-used in all the iterations, is expected to reside on cache memory, and the memory-operation cost for the gate matrix is counted at once. Thus, the number of memory operation is effectively $O(2^{n} + 2^{2m})$. Typically, because we consider the case when $m \ll n$, the memory-operation cost is further approximated as $O(2^n)$.

Although any basic gate can be treated as a dense matrix gate, quantum gates in quantum computing research sometimes have an additional structure in a gate matrix $\tilde{K}$. By utilizing this structure, the arithmetic-operation cost, the memory-operation cost, or both can be decreased. This motivates us to define specialized subclasses of dense matrix gates for quantum gates with structured gate matrices. Here, we show functions for several types of basic gates with a structure. Note that the relation between arithmetic- and memory-operation costs and the total computation time is discussed in Sec.\,\ref{sec:optimization}.

Let $M_c$ be a subset of target qubits and $c$~($0 \le c < 2^{|M_c|}$) be an integer, where $|\cdot|$ represents the number of elements in a given set. Any gate matrix $\tilde{K}$ can be represented in the form $\tilde{K} = \sum_{x,y=0}^{2^{|M_c|}-1} \ket{x}\bra{y} \otimes L_{x,y}$, where the first part of the tensor product acts on the space of qubits in $M_c$, and the latter part acts on the space of target qubits except for $M_c$. 
Suppose a gate matrix $\tilde{K}$ which satisfies $L_{x,y}=0$ if $x\neq y$ and $L_{x,x}=I$ if $x \neq c$. A quantum gate with such a gate matrix $\tilde{K}$ is called controlled quantum gates. We can describe a gate matrix of a controlled quantum gate with an integer $c$ and complex matrix $L_{c,c}$. Let $m_c := |M_c|$ be the number of control qubits, and $m_t := m-m_c$ be the number of qubits on which $L_{c,c}$ act. Then, this gate can be applied with arithmetic-operation costs $O(2^{n-m_c+m_t})$ and memory-operation costs $O(2^{n-m_c})$. 
In Qulacs, we can specify the pair of the digit of control index $c$ and corresponding item in $M_c$ with a member function \texttt{add\_control\_qubit}. 

When a gate matrix has a small number of non-zero elements, it is called sparse. When $\tilde{K}$ is a sparse matrix, the arithmetic- and memory-operation costs can be decreased according to the number of non-zero elements. A quantum gate with a sparse gate matrix can be generated with  \texttt{SparseMatrix}.

Another special case is a diagonal matrix, which is a matrix with non-zero elements only in the diagonal elements. In this case, arithmetic- and memory-operation costs become $O(2^n)$. These costs are independent of the number of target qubits $m$. In this case, \texttt{DiagonalMatrix} can be used for generating diagonal matrix gates.

An action of reversible Boolean functions in classical computing can always be represented as a permutation matrix, which is a matrix where there is a single unity element in each row and column. 
\texttt{ReversibleBoolean} creates a unitary operation with a permutation matrix by supplying a function that returns the index of a column with a unity element from the index of a given row. This function is applicable not only for reversible circuits but also for creating and annihilating operators as a product of a permutation matrix and a diagonal matrix. Its arithmetic- and memory-operation costs are $O(2^n)$ and are also independent of the number of target qubits $m$.

A set of $m$-qubit Pauli matrices is defined as a tensor product of Pauli matrices $\{I, X, Y, Z\}^{\otimes m}$, where 
\begin{align}
I &= \left(\begin{matrix}1 & 0 \\ 0 & 1 \end{matrix}\right), X = \left(\begin{matrix}0 & 1 \\ 1 & 0 \end{matrix}\right), \nonumber \\
Y &= \left(\begin{matrix} 0 & -i \\ i & 0 \end{matrix}\right), Z = \left(\begin{matrix} 1 & 0 \\ 0 & -1 \end{matrix}\right).
\end{align}
A Pauli gate is a gate in which the gate matrix is a Pauli matrix. 
By assigning numbers $\{0,1,2,3\}$ to Pauli matrices $\{I, X, Y, Z\}$, respectively, a Pauli matrix can be represented with a sequence of integers $\{0,1,2,3\}^m$. The \texttt{Pauli} function generates a basic gate with a Pauli matrix represented by a sequence of assigned integers. Its arithmetic- and memory-operation costs are $O(2^n)$ and are independent of $m$. 

Since a set of $m$-qubit Pauli matrices is a basis of $2^m \times 2^m$ matrices, any $2^m \times 2^m$ matrix can be represented as a linear combination of $m$-qubit Pauli matrices. Furthermore, any self-adjoint matrix can be represented as a linear combination of $m$-qubit Pauli matrices with real coefficients. Therefore, any unitary gate matrix can be represented in the form $\tilde{K} = \exp(i \sum_P \theta_P P)$, where $\theta_P$ is a real coefficient. Suppose a quantum gate such that $\theta_P=0$ if $P \neq Q$, where $Q$ is a certain Pauli matrix. Such a quantum gate is called a Pauli rotation gate. 
\texttt{PauliRotation} can be used for generating a Pauli rotation gate with a description of Pauli matrix $Q$ and rotation angle $\theta_Q$. 
Its arithmetic- and memory-operation costs are also $O(2^n)$. 
Note that quantum gates with multiple non-zero rotation angles, which are vital for simulating the dynamics of quantum systems under a given Hamiltonian, can be generated with \texttt{DenseMatrix} function with an explicit matrix representation of $\exp(i \sum_P \theta_P P)$ or generated as a Trotterized quantum circuit with observable. For the latter, see Sec.\,\ref{subsec:observable}.
Listing.\,\ref{alg_special} shows examples. Qulacs has several other specializations for basic gates, which are detailed in the online manuals~\cite{Qulacs}.

\begin{lstlisting}[language=Python, caption=An example Python program that applies several basic gates to quantum states., label=alg_special]
import numpy as np
from scipy.sparse import csr_matrix
from qulacs import StateVector
from qulacs.gate import DenseMatrix, SparseMatrix, DiagonalMatrix, Pauli, PauliRotation, ReversibleBoolean
state = StateVector(10)

# Update a quantum state with a controlled dense matrix gate
target_list = [1]
gate_matrix = [[0, 1],[1, 0]]
control_gate = DenseMatrix(target_list, gate_matrix)
## Act when the 2-nd qubit is |0>
control_gate.add_control_qubit(2, 0)
## Act when the 3-rd qubit is |1>
control_gate.add_control_qubit(3, 1)
control_gate.update_quantum_state(state)

# Update a quantum state with a sparse matrix gate
target_list = [2, 1]
sparse_matrix = csr_matrix(
    ([1,1], ([0,3], [0,3])), 
    shape=(4,4), dtype=complex)
sparse_gate = SparseMatrix(target_list, sparse_matrix)
sparse_gate.update_quantum_state(state)

# update a quantum state with a diagonal matrix gate
target_list = [3, 5]
diagonal_element = [1, -1, -1, 1]
diagonal_gate = DiagonalMatrix(target_list, diagonal_element)
diagonal_gate.update_quantum_state(state)

# Update a quantum state with a permutation matrix gate
def basis_to_basis(index, dim):
  return (index+3)%dim

target_list = [0, 3, 4]
rev_gate = ReversibleBoolean(target_list, basis_to_basis)
rev_gate.update_quantum_state(state)

# Update a quantum state with Pauli gate
target_list = [1,2]
pauli_ids = [3,2]
pauli_gate = Pauli(target_list, pauli_ids)
pauli_gate.update_quantum_state(state)

# Update a quantum state with a Pauli rotation gate
target_list = [1,2]
pauli_ids = [3,2]
rotation_angle = np.pi/5
rot_gate = PauliRotation(target_list, pauli_ids, rotation_angle)
rot_gate.update_quantum_state(state)
\end{lstlisting}

\subsubsection{Quantum map}
Quantum maps are a general operation, which includes all the quantum maps that cannot be represented as basic gates such as measurement, noisy operation, and feedback operation.

The most general form of physical operations without measurements is a completely-positive trace-preserving (CPTP)~\cite{nielsen2002quantum}. According to the operator-sum representation, this map can be represented as $\rho \mapsto \sum_i K_i \rho K_i^{\dagger}$, where $\rho$ is the density matrix and $K_i$ is called the Kraus operator. The map must satisfy the condition $\sum_i K_i^{\dagger} K_i = I$. 
In Qulacs, a list of basic gates, which represent the action of Kraus operators $\{K_i\}$, is required to create a CPTP-map.
When a density matrix is used as a representation of quantum states, $\rho$ is mapped to $\sum_i K_i \rho K_i^{\dagger}$. On the other hand, when a state vector is used as a representation of quantum states, the $i$-th Kraus operator is chosen with probability $p_i = |K_i \ket{\psi}|^2$. Then, the state vector is mapped to $\cfrac{K_i \ket{\psi}}{\sqrt{p_i}}$. Suppose that the number of Kraus operators is $k$ and each Kraus operator acts on $m$-qubits, then in the worst case, the arithmetic- and memory-operation costs become $O(2^{n+m}k)$. The CPTP-map can be created with a \texttt{CPTP} function.

One of the most general representations of all physically achievable operations, including measurements, is the completely-positive (CP) instrument. In Qulacs, this operation is the same as a CPTP-map except that the index of the chosen Kraus operator is stored in the classical register of \texttt{QuantumStateBase} and can be used later. This map can be generated with a function \texttt{Instrument}. The arithmetic- and memory-operation costs are the same as CPTP-map.

A CPTP-map is called unital when it maps a maximally mixed state to itself. A unital CPTP-map can be represented as a probabilistic application of unitary operations (i.e., for all $i$, $K_i$ has a form  $K_i = \sqrt{p_i} U_i$, where $p_i$ is a real value and $U_i$ is unitary). Unlike a CPTP-map, the arithmetic- and memory-operation costs of unital maps decrease to $O(2^{n+m})$. The costs become independent of the number of Kraus operators since the probability distribution for sampling a Kraus operator is independent of the input state and can be given in advance. A unital CPTP-map can generated with \texttt{Probabilistic} function.

An adaptive map is one that acts on the quantum states only when a given classical condition is satisfied. This map requires a Boolean function that determines an output according to the classical registers. Then, a map is applied only when the returned value of the Boolean function is \texttt{True}. This map is useful for treating feedback and feedforward operations such as the heralded operation, readout initialization, measurement-based quantum computation, and look-up table decoder for quantum error correction. This gate can be generated with a function named \texttt{Adaptive}. Its arithmetic- and memory-operation costs are dependent on a given gate and the probability that the given condition is satisfied.

Listing.\,\ref{alg_general} shows the example codes of these general maps. There are several other forms of general gates in Qulacs for a specific research purpose. See the online manual~\cite{Qulacs} for details.

\begin{lstlisting}[language=Python, caption=An example Python program that applies general quantum maps to quantum states., label=alg_general]
from qulacs import StateVector
from qulacs.gate import X, Y, Z, P0, P1
from qulacs.gate import Instrument, CPTP, Probabilistic, Adaptive

state = StateVector(3)
gate_list = [X(0), Y(0), Z(0)]


# Update a quantum state with a CPTP map
gate_list = [P0(0), P1(0)]
cptp_gate = CPTP(gate_list)
cptp_gate.update_quantum_state(state)

# Update a quantum state with a CP instrument
classical_register = 0
gate_list = [P0(0), P1(0)]
inst_gate = Instrument(gate_list, classical_register)
inst_gate.update_quantum_state(state)

# Get and set values in the classical register
value = state.get_classical_value(classical_register)
state.set_classical_value(classical_register, 1-value)

# Update a quantum state with a unital gate
gate_list = [X(0), Y(0), Z(0)]
prob_list = [0.2, 0.3, 0.1]
prob_gate = Probabilistic(prob_list, gate_list)
prob_gate.update_quantum_state(state)

# Update a quantum state with an adaptive gate
def func(classical_register_list):
    return classical_register_list[0] == 0

gate = X(0)
adap_gate = Adaptive(gate, condition=func)
adap_gate.update_quantum_state(state)
\end{lstlisting}

\subsubsection{Named gate}
Since quantum gates with several specific gate matrices and Kraus operators are frequently used in quantum computing research, functions to generate these gates are defined. TABLE.\ref{Tab:named_gate} lists these named gates. Although some of these function calls are simply redirected to the definition as quantum maps, the following gates are redirected optimized update functions: \texttt{X,Y,Z,H,CNOT,SWAP,CZ}. Thus, when these quantum gates are used, they should be instantiated using these functions. 

\begin{table*}[t!]
\centering
\begin{tabular}{c|c|l}
Category & Name & Description  \\ \hline 
\hline
Single-qubit gate & \texttt{X} & Pauli-$X$ gate \\ 
& \texttt{Y} & Pauli-$Y$ gate \\ 
& \texttt{Z} & Pauli-$Z$ gate \\ 
& \texttt{sqrtX}    & $\pi/4$ rotation of Pauli-$X$ gate \\ 
& \texttt{sqrtXdag} & $-\pi/4$ rotation of Pauli-$X$ gate \\ 
& \texttt{sqrtY}    & $\pi/4$ rotation of Pauli-$Y$ gate \\ 
& \texttt{sqrtYdag} & $-\pi/4$ rotation of Pauli-$Y$ gate \\ 
& \texttt{S} & $\pi/4$ rotation of Pauli-$Z$ gate \\ 
& \texttt{Sdag} & $-\pi/4$ rotation of Pauli-$Z$ gate \\ 
& \texttt{T} & $\pi/8$ rotation of Pauli-$Z$ gate \\ 
& \texttt{Tdag} & $-\pi/8$ rotation of Pauli-$Z$ gate \\ 
& \texttt{H} & Hadamard gate \\ 
\hline
Two-qubit gate & \texttt{CNOT} & Controlled-NOT gate \\ 
& \texttt{CZ}   & Controlled-Z gate \\ 
& \texttt{SWAP} & SWAP gate \\ 
\hline
Three-qubit gate & \texttt{TOFFOLI} & TOFFOLI gate \\ 
& \texttt{FREDKIN} & FREDKIN gate \\ 
\hline
Single-qubit & \texttt{RX} & Pauli-$X$ rotation: $\exp(i\theta X/2)$ \\ 
rotation gate & \texttt{RY} & Pauli-$Y$ rotation: $\exp(i\theta Y/2)$ \\ 
& \texttt{RZ} & Pauli-$Z$ rotation: $\exp(i\theta Z/2)$ \\ 
& \texttt{U1} & Rotate phase of LO~(Local oscillator) \\ 
& \texttt{U2} & Rotate phase of LO with single $\pi/2$-pulse \\ 
& \texttt{U3} & Rotate phase of LO with two $\pi/2$-pulses \\ 
\hline
Projection and & \texttt{P0} & Projection matrix to $\ket{0}$ state \\ 
measurement & \texttt{P1} & Projection matrix to $\ket{1}$ state \\ 
& \texttt{Measurement} & Single qubit measurement with $Z$-basis\\ 
\hline
Noise & \texttt{BitFlipNoise}      & Probabilistic Pauli-$X$ operation \\ 
& \texttt{DephasingNoise}    & Probabilistic Pauli-$Z$ operation \\ 
& \texttt{DepolarizingNoise} & Single-qubit uniform depolarizing noise \\ 
& \texttt{TwoQubitDepolarizingNoise} & Two-qubit uniform depolarizing noise \\ 
& \texttt{AmplitudeDampingNoise} & Single-qubit amplitude damping noise \\ 
\end{tabular}
\caption{Partial listing of named gates in Qulacs. These are all defined in the \texttt{qulacs.gate} module. Several gates have optimized functions, while others are alias to quantum gates. For the definition of \texttt{U1}, \texttt{U2}, and \texttt{U3}, see the reference of IBMQ~\cite{cross2017open}}
\label{Tab:named_gate}
\end{table*}

\subsection{Quantum circuit}
In Qulacs, a quantum circuit is represented as a simple array of quantum gates. The \texttt{QuantumCircuit} class instantiates quantum circuits. The \texttt{add\_gate} function inserts a quantum gate to a circuit with a given position. If a position is not given, the gate is appended to the last of a quantum circuit. The \texttt{remove\_gate} function removes a quantum gate at a given position. 
By calling a member function \texttt{update\_quantum\_state}, all the contained quantum gates are applied to a given state sequentially. 

When users need to treat parametric quantum gates and circuits, the \texttt{ParametricQuantumCircuit} class should be used, which provides functions to treat parameters in quantum circuits. By adding \texttt{ParametricRX}, \texttt{ParametricRY}, \texttt{ParametricRZ}, and \texttt{PrametricPauliRotation} gates with the \texttt{add\_parametric\_gate} function, their rotation angles can be set and varied with the \texttt{get\_parameter} and \texttt{set\_parameter} functions. Listing.\,\ref{alg_circuit} shows some examples.

\begin{lstlisting}[language=Python, caption=An example Python program that generates quantum circuits and parametric ones., label=alg_circuit]
from qulacs import StateVector, QuantumCircuit, ParametricQuantumCircuit
from qulacs.gate import DenseMatrix, H, CNOT
from qulacs.gate import ParametricRX, ParametricRY, ParametricPauliRotation

# Create a quantum circuit and add quantum gates
n = 5
circuit = QuantumCircuit(n)
circuit.add_gate(DenseMatrix([1], [[0,1],[1,0]]))
for index in range(n):
  circuit.add_gate(H(index))

# Insert quantum gates into a given position
position = 1
circuit.add_gate(CNOT(2,3), position)

# Remove a quantum gate at a position
position = 0
circuit.remove_gate(position)

# Compute the depth of the quantum circuit
depth = circuit.calculate_depth()

# Get the number of quantum gates
gate_count = circuit.get_gate_count()

# Get a copy of the quantum gate at a position
position = 0
gate = circuit.get_gate(position)

# Update a state vector with the quantum circuit
state = StateVector(n)
circuit.update_quantum_state(state)

# Create parametric quantum circuit
par_circuit = ParametricQuantumCircuit(n)
par_circuit.add_parametric_gate(ParametricRX(0, 0.1))
par_circuit.add_parametric_gate(ParametricRY(1, 0.1))
par_circuit.add_parametric_gate(ParametricPauliRotation([0,1], [1,1], 0.1))

# Get the number of parameters in the quantum circuit
par_count = par_circuit.get_parameter_count()

# Get and set a parameter at a position
index = 0
angle = 0.2
value = par_circuit.get_parameter(index)
par_circuit.set_parameter(index, angle)

# Get the position of a parametric gate from the index of a parameter
position = par_circuit.get_parametric_gate_position(index)
\end{lstlisting}

\begin{widetext}
\subsection{Observable}
\label{subsec:observable}
In quantum physics, physical values are obtained as an expectation value of a self-adjoint operator named observable. In Qulacs, any observable $O$ is represented as a linear combination of Pauli matrices with real coefficients, i.e., $O = \sum_P \alpha_P P$ where $\alpha_P \in \mathbb{R}$. A Pauli term in observable $\alpha_P P$ is constructed with the \texttt{PauliOperator} class. 
An observable is generated with the \texttt{Observable} class. The \texttt{add\_operator} function adds a Pauli term to an observable. 
Then, the \texttt{get\_expectation\_value} function computes the expectation value of an observable according to a given quantum state, and the \texttt{get\_transition\_amplitude} function computes the transition amplitude of an observable according to two quantum states. 

A Hamiltonian is an observable for the energy of a quantum system, and a unitary operator for time evolution is described as an exponential of the Hamiltonian operator with an imaginary coefficient. 
The \texttt{add\_observable\_rotation\_gate} function adds a set of quantum gates for simulating the time evolution under a given Hamiltonian, which is generated with the Trotter decomposition, to a quantum circuit.

Expectation values or transition amplitudes of Hamiltonian of molecules are frequently studied in the field of NISQ applications~\cite{tamiya2020calculating,ibe2020calculating,Kawai_2020}. They are usually represented as a linear combination of products of fermionic operators. They can be converted to an observable with Pauli operators using the Jordan-Wigner transformation~\cite{jordan1928pauli} or the Bravyi-Kitaev transformation~\cite{bravyi2002fermionic}, which are implemented in OpenFermion~\cite{mcclean2020openfermion}. The format of OpenFermion is loaded with the \texttt{create\_quantum\_operator\_from\_openfermion\_text} function, which allows the output of OpenFermion to be interpreted as a format of Qulacs.
The \texttt{GeneralQuantumOperator} class can be used to generate observables that are not self-adjoint. Listing.\,\ref{alg_obs} shows example codes for treating observables and evaluating expectation values.
\begin{lstlisting}[language=Python, caption=An example Python program that generates and evaluates observables., label=alg_obs]
from qulacs import Observable, PauliOperator, StateVector, QuantumCircuit
from qulacs.quantum_operator import create_quantum_operator_from_openfermion_text

# Construct a Pauli operator
coef = 2.0
Pauli_string = "X 0 X 1 Y 2 Z 4"
pauli = PauliOperator(Pauli_string, coef)

# Create an observable acting on n qubits
n = 5
observable = Observable(n)
# Add a Pauli operator to the observable
observable.add_operator(pauli)
# or directly add it with coef and str
observable.add_operator(0.5, "Y 1 Z 4")

# Get the number of terms in the observable
term_count = observable.get_term_count()

# Get the number of qubit on which the observable acts
qubit_count = observable.get_qubit_count()

# Get a specific term as PauliOperator
index = 1
pauli = observable.get_term(index)

# Calculate the expectation value <a|H|a>
state = StateVector(n)
state.set_Haar_random_state(0)
expect = observable.get_expectation_value(state)

# Calculate the transition amplitude <a|H|b>
bra = StateVector(n)
bra.set_Haar_random_state(1)
trans_amp = observable.get_transition_amplitude(bra, state)

# Create a quantum circuit to simulate
#  the time evolution by a given observable
# Observable is Trotterized with given slice count.
circuit = QuantumCircuit(n)
angle = 0.1
t_slice = 100
circuit.add_observable_rotation_gate(obs, angle, t_slice)
circuit.update_quantum_state(state)

# Load an observable from OpenFermion text
open_fermion_text = """
(-0.8126100000000005+0j) [] +
(0.04532175+0j) [X0 Z1 X2] +
(0.04532175+0j) [X0 Z1 X2 Z3] +
(0.04532175+0j) [Y0 Z1 Y2] +
(0.04532175+0j) [Y0 Z1 Y2 Z3] +
(0.17120100000000002+0j) [Z0] +
(0.17120100000000002+0j) [Z0 Z1] +
(0.165868+0j) [Z0 Z1 Z2] +
(0.165868+0j) [Z0 Z1 Z2 Z3] +
(0.12054625+0j) [Z0 Z2] +
(0.12054625+0j) [Z0 Z2 Z3] +
(0.16862325+0j) [Z1] +
(-0.22279649999999998+0j) [Z1 Z2 Z3] +
(0.17434925+0j) [Z1 Z3] +
(-0.22279649999999998+0j) [Z2]
"""
obs_of = create_quantum_operator_from_openfermion_text(open_fermion_text)
\end{lstlisting}
\end{widetext}

\section{Optimizations}
\label{sec:optimization}
In this section, we discuss possible performance bottlenecks in quantum simulation, and discuss several optimization techniques for different computing devices.

\subsection{Background}
Since an application of a quantum gate is a large number of simple iterations, a time for circuit simulation can be roughly estimated as the sum of constant-time overheads for invoking functions and times for processing arithmetic and memory operations. 
Several factors such as an overhead to call C++ function via python interfaces, functions with parallelization, and GPU kernels incur additional overheads in computation. These additional overheads are quantitatively discussed later in Sec.\,\ref{sec:performance}. 
The times for processing arithmetic and memory operations are determined by the number of operations divided by throughput. For the time for arithmetic operations, the number of operations that can be processed in a unit second is vital, which is known as floating-point operations per second (FLOPS). To process complex numbers in the CPU, we need to load complex numbers representing quantum states from the CPU cache or main RAM to the CPU registers. The size of data per unit time that we can transfer between a memory and processor is called the bandwidth of the memory.
Let the time for the additional overheads be $T_{\rm over}$, the number of arithmetic operations be $N_{\rm com}$, the number of memory operations be $N_{\rm mem}$, the FLOPS of CPU be $V_{\rm FLOPS}$, and the memory bandwidth, i.e., the number of complex numbers which we can transfer, be $V_{\rm BW}$. $T_{\rm over}$ is determined by a design of a library, $N_{\rm mem}$ and $N_{\rm com}$ are determined by a quantum gate to apply, and $V_{\rm FLOPS}$ and $V_{\rm BW}$ are determined by a computing device.
Then, an approximate total time for applying a quantum gate $T_{\rm gate}$ is lower bounded as 
\begin{align}
T_{\rm gate} \geq T_{\rm over} + \max (N_{\rm com}/V_{\rm FLOPS}, N_{\rm mem}/V_{\rm BW})
\end{align}
when computation and memory operations do not block each other. While actual processing is not necessarily simplified to this equation, estimation with this equation works well when we develop a quantum circuit simulator.

To develop a quantum circuit simulator satisfying the demands shown in Sec.\,\ref{sec:expected_use}, we can find several basic directions from this equation. In the case of a small number of qubits, $T_{\rm over}$ becomes a dominant factor. Thus, the pre- and post-processing for applying quantum gates should be minimized. In Qulacs, every core function is designed to minimize overheads.
When the number of qubits $n$ increases, values of $N_{\rm com}/V_{\rm FLOPS}$ and $N_{\rm mem}/V_{\rm BW}$ grow exponentially to $n$, and they become larger than the overhead $T_{\rm over}$. In this region, the values $N_{\rm com}/V_{\rm FLOPS}$ and $N_{\rm mem}/V_{\rm BW}$ should be minimized. If there are two ways to update quantum states, and if one has smaller $N_{\rm com}$ and $N_{\rm mem}$ than the other, the first one should be chosen to minimize $T_{\rm gate}$. To this end, Qulacs provides several specialized update functions that utilize the structure of gate matrices as introduced in Sec.\,\ref{sec:implementation}.
In this section, we show four additional techniques to minimize $T_{\rm gate}$ when $N_{\rm com}/V_{\rm FLOPS}$ and $N_{\rm mem}/V_{\rm BW}$ are dominant: SIMD Optimization, multi-threading with OpenMP, quantum circuit optimization, and GPU acceleration.

\subsection{SIMD optimization}
Recent processors support SIMD~(single-instruction, multiple data) instructions, which can apply the same operation to multiple data simultaneously. Qulacs utilizes instructions named Intel AVX2~\cite{AVX2}, in which up to 256-bit data can be processed simultaneously. 
When a quantum state is represented as an array of double-precision real values, we can load, store, and process four real values (i.e. two complex numbers) simultaneously. Thus, the use of AVX2 can reduce the number of instructions $N_{\rm com}$ by a factor of four at most. 
In Qulacs, several update functions are optimized with AVX2 instructions by hand. When Qulacs is being installed, the installer checks if a system supports such a feature. If it is supported, the library is built with AVX2 instructions enabled.

Here, we discuss our SIMD optimization techniques using dense matrix gates. However, it is worth noting that our techniques are applicable to the other quantum gates. The naive implementation of dense matrix gates is shown in Listing.\,\ref{alg_pseudo}. 
We implemented two SIMD versions of the $B_1$-loop. When all the indices of the target qubits are large, it is not possible to SIMDize it as it is because the state vector elements required in one iteration of the $B_0$-loop are scattered across non-contiguous memory locations. However, since the value of the adjacent memory location is always loaded in the next iteration, we unroll the $B_0$-loop according to the number of target qubits to enable AVX2's SIMD load/store operations. On the other hand, when there is a target qubit with a small index, we can SIMDize it without such an unrolling because the required state vector elements are already adjacent. 
Note that the overhead of enumerating $B_0$ and $B_1$ is not negligible when the number of target qubits $m$ is small. We reduce the cost of the enumeration of $B_0$ and $B_1$ as follows: Instead of listing all the items in $B_0$ beforehand, the $i$-th element of $B_0$ is computed from the index $i$ in each iteration using bit-wise operation techniques. In contrast, the list of $B_1$ is computed before the $B_0$-loop since the size of $B_1$ is typically small. This implementation is useful for parallelizing iterations with multi-threading by OpenMP, which is explained in the Sec.\,\ref{sec:OpenMP}. When a gate matrix has a structure and basic gates other than a dense matrix can be utilized, an updated state vector can be calculated without matrix-vector multiplication, and thus a different optimization is applied.

\subsection{Multi-threading with OpenMP}
\label{sec:OpenMP}
Since recent CPUs contain multiple processing cores, executing iterations in update and evaluation functions in parallel can increase the effective instruction throughput $V_{\rm FLOPS}$. 
The use of multiple cores is effective particularly when FLOPS is a performance bottleneck (compute-bound), and each core has a certain amount of workload.
Qulacs parallelizes the execution of update functions using OpenMP directives~\cite{OpenMP}. The number of threads used in these functions can be controlled with environment variable \texttt{OMP\_NUM\_THREADS}.
The naive implementation shown in Listing.\,\ref{alg_pseudo} consists of two loops: $B_0$-loop and $B_1$-loop. In Qulacs, the $B_0$-loop is parallelized to maximize the amount of workload for each core and to minimize the overhead incurred by multi-threading.
Specifically, the parallelized loop iterates over $[0..2^{n-m}-1]$ and the iteration space is chunked into $T$ chunks, where $T$ is the number of threads. In the loop body, the $i$-th element of $B_0$ is computed on-the-fly from the loop index, which enables the even distribution of workload across threads. In contrast, the list of $B_1$ is created before executing the parallel loop.
While the data of $B_1$ is accessed from every thread, its overhead is expected to be not so high because $B_1$ is not updated during the loop and its size $|B_1|=2^m$ is typically small enough to store within CPU registers or caches. A buffer for a temporal state vector (\texttt{temp\_vector} in Listing.\,\ref{alg_pseudo}) is a thread-local array that can be read and written by each thread independently. Thus, a buffer space for $T \times 2^m$ data is allocated before the $B_0$-loop, so that each $2^m$ block can be used by a corresponding thread, and is deallocated at the end of the parallel loop.
Note that when $n$ and $m$ are small, the amount of workload for each thread becomes small. In that case, the overhead due to multi-threading becomes larger than the speed-up by multi-threading. Therefore, Qulacs automatically disables multi-threading if the number of qubits $n$ is smaller than a threshold value even when \texttt{OMP\_NUM\_THREADS} is set to 2 or more. This threshold value is empirically determined according to the number of $m$ and a type of basic gates.

\subsection{Circuit optimization}
When the SIMD and multi-threading with OpenMP are enabled, a time for processing arithmetic operations $N_{\rm com}/V_{\rm FLOPS}$ becomes smaller than that for processing memory operations $N_{\rm mem}/V_{\rm BW}$, and a computing time $T_{\rm gate}$ is devoted to transferring data between the main RAM or CPU cache and the CPU. Circuit optimization is a technique to trade $N_{\rm mem}$ and $N_{\rm com}$, i.e., we can reduce $N_{\rm mem}$ by sacrificing $N_{\rm com}$ with a circuit optimization technique.
For simplicity, we suppose the case when we are given a quantum circuit that consists only of dense matrix gates. As we discussed in Sec.\,\ref{sec:implementation}, the number of arithmetic operations for a dense matrix gate acting on a set of qubits $M$ increases as $2^{n+|M|}$ and that of memory operations as $2^n$. 
We suppose a situation where we need to apply two successive dense matrix gates which act on a set of qubits $M_1$ and $M_2$, where $|M_2| \le |M_1| \ll n$. Here, we can merge these two quantum gates to a single quantum gate and apply it to a quantum state instead of applying two gates one by one. Then, the number of arithmetic operations changes from $2^{n+|M_1|} + 2^{n+|M_2|}$ to $2^{n+|M_1 \cup M_2|}$, and the number of memory operations changes from $2^{n+1}$ to $2^{n}$. Note that while there is a cost for computing a gate matrix of a merged quantum gate, the complexity for computing the gate matrix is at most $O(2^{3|M_1 \cup M_2|})$, which is negligible compared with the cost for applying quantum gates to quantum states when $|M_1 \cup M_2| \ll n$.
Thus, we can halve the number of memory operations by multiplying that of arithmetic operations by $2^{|M_1 \cup M_2| - |M_1|}$.
This means when $M_2$ is a subset of $M_1$, we can decrease the number of memory operations with a negligible penalty. Even if there is a penalty, we should perform merge operations until $N_{\rm mem}/V_{\rm BW}$ is balanced to $N_{\rm com}/V_{\rm FLOPS}$. The optimal size of merged quantum gates is relevant to a value of $V_{\rm BW}$ divided by $V_{\rm FLOPS}$, which is called a BF ratio. 
Although the BF ratio varies depending on a chosen CPU and memory and the best strategy is dependent on the structure of given quantum circuits, it is typically optimal to merge quantum gates until every quantum gate acts on at most two qubits in the case of random quantum circuits.

To minimize the time for simulating quantum circuits with this technique, Qulacs provides the \texttt{gate.merge} function to create a merged quantum gate from two basic gates. The \texttt{merge\_all} function of \texttt{QuantumCircuitOptimizer} class merges all the basic quantum gates in a given quantum circuit to a single gate. While Qulacs expects that this kind of optimization should be performed by the user according to the tasks, Qulacs provides two strategies to quickly merge several quantum gates in quantum circuits: light and heavy optimizations.
The light optimization finds a pair of gates such that they are neighboring and target qubits of one gate is a subset of the other with a greedy algorithm and merge it. As discussed, such a pair of gates can be merged without any penalty. 
By contrast, the heavy optimization merges two quantum gates when the following conditions are satisfied; two gates can be moved to neighboring positions by repetitively swap the commutative quantum gates, and the number of target qubits of the merged quantum gate is not larger than the given block size. While the heavy optimization can decrease the number of quantum gates compared to the light optimization, it consumes a longer time for optimization. When we measure a time for optimization as a part of the simulation time, overall performance benefits from the heavy optimization varies depending on the structure of quantum circuits and computing devices. 
Note that every quantum gate keeps the information about commutable Pauli-basis for each qubit index. For example, a CNOT gate can commute with the Pauli-$Z$ basis at the controlled qubit and with the Pauli-$X$ basis at the target qubit. By utilizing this information, the heavy optimization can check whether two quantum gates can commute or not quickly. Note that several quantum gates such as CPTP-maps and parametric quantum gates cannot be merged in the optimization process. 
Listing.\,\ref{alg_opt_circuit} shows example codes for circuit optimization.
\begin{lstlisting}[language=Python, caption=An example Python program that performs circuit optimization., label=alg_opt_circuit]
from qulacs import StateVector, QuantumCircuit
from qulacs.gate import RandomUnitary, CNOT, merge
from qulacs.circuit import QuantumCircuitOptimizer
n = 4
gate1 = RandomUnitary([0,1])
gate2 = RandomUnitary([2,1])

# Create a merged gate 
merged_gate = merge(gate1, gate2)


# Create an example circuit
layer_count = 5
circuit = QuantumCircuit(n)
for layer_index in range(layer_count):
  for index in range(n):
    circuit.add_gate(RandomUnitary([index]))
  for index in range(layer_index%2, n-1, 2):
    circuit.add_gate(CNOT(index, index+1))

for index in range(n):
  circuit.add_gate(RandomUnitary([index]))

# Optimize the quantum circuit
optimizer = QuantumCircuitOptimizer()
## Merge all the gates to a single unitary
whole_unitary = optimizer.merge_all(circuit)
## Light optimization
circuit_opt1 = circuit.copy()
optimizer.optimize_light(circuit_opt1)
## Heavy optimization
circuit_opt2 = circuit.copy()
optimizer.optimize(circuit_opt2, block_size=3)
\end{lstlisting}

\subsection{GPU acceleration}
A GPU is a computing device that has a higher memory bandwidth than a CPU, and it also has higher peak performance than a CPU since a GPU has a large number of processing cores. Therefore, it is possible that a quantum simulation on a GPU is significantly faster than that on a CPU in certain cases.
Also, since a GPU has several types of memories with different performance characteristics, it is important to consider how to access and where to place state vectors and gate matrices for achieving the performance close to the peak FLOPS. Here, we discuss optimization techniques for NVIDIA GPUs, However, our techniques should apply to other GPUs such as AMD GPUs.
In NVIDIA GPUs, there are six types of memories in a GPU: registers, shared memory, local memory, constant memory, texture memory, and global memory. Qulacs uses registers, shared memory, constant memory, and global memory for calculation.
The global memory has the largest capacity but has the highest latency and lowest bandwidth in the memories. 
By contrast, the registers can be accessed with the lowest latency in the memories but their capacity is limited. 
The shared memory is a memory that is shared by and synchronized among threads in a block. This memory has a larger capacity than the registers and has higher bandwidth and lower latency than the global memory. 
The constant memory is a memory that is read-only for GPU kernels and writable from a CPU host. Since memory accesses to the constant memory are cached, we can use the constant memory as a read-only memory of which the bandwidth and latency are almost the same as those of the registers and the capacity is larger than the registers.
In our implementation, a whole state vector is allocated in the global memory since the size of the other memories is typically not sufficient for storing the state vector. 
In each iteration, a gate matrix, which consists of $4^m$ complex numbers, and a temporal state vector, which consists of $2^m$ complex numbers of the whole state vector (i.e., a variable \texttt{temp\_vector} in Listing.\,\ref{alg_pseudo}), should be temporally stored in a high-bandwidth memory to minimize the time for memory operations. To this end, in Qulacs, memories used for storing a gate matrix and a temporal state vector is chosen according to the number of target qubits.
The memory used for storing a temporal state vector is chosen as follows:
When the number of target qubits is no more than four, we expect that a temporal state vector is fetched from the global memory to registers and arithmetic operations are performed in each thread. 
When the number of target qubits is between five to eleven, a temporal state vector is loaded to the shared memory since it must be synchronized in the block. 
Otherwise, a temporal state vector is allocated on the global memory, and matrix-vector multiplication is performed with that.
The memory used for storing a gate matrix is chosen as follows:
If the number of the target qubits is no more than three, it tends to be placed in registers by the compiler by giving elements of the gate matrix as arguments of the function call.
When the number of target qubits is four or five, we use constant memory since a gate matrix is common and constant for all the threads.
When the number of qubits is six or more, gate matrices are stored in the global memory.

A state vector can be allocated on a GPU with the \texttt{StateVectorGpu} class. When the computing node has multiple GPUs, the index of GPU to allocate state vectors can be identified with the second argument of the constructor function. Once a state vector is allocated on a GPU, every processing on it is performed with the selected GPU unless a state vector is converted to \texttt{StateVector}. Hence, multiple tasks can be simulated with multiple GPUs simultaneously.
Listing.\,\ref{alg_obs} shows example codes for GPU computing.
\begin{lstlisting}[language=Python, caption=An example Python program that creates and manipulates quantum states within a GPU., label=alg_gpu]
from qulacs import StateVectorGpu, StateVector
from qulacs.gate import H, CNOT

# Create a state vector within the 0-th GPU.
n = 10
gpu_id = 0
state = StateVectorGpu(n, gpu_id)

# Apply Hadamard and CNOT gates to StateVector on GPU
H(0).update_quantum_state(state)
CNOT(0,1).update_quantum_state(state)

# Load the state vector on GPU to main RAM
state_cpu = StateVector(n)
state_cpu.load(state)

# Get the state vector as numpy array
numpy_array = state.get_vector()
\end{lstlisting}

\section{Numerical performance}
\label{sec:performance}
Here, we compare the simulation times for several computing tasks with Qulacs using various types of settings. 
For the benchmark, we use Qulacs~0.2.0 with Python 3.8.5 on Ubuntu 20.04.1 LTS. Qulacs is is compiled with GNU Compiler Collection (GCC) 9.3.0 and with the options: \texttt{-O2 -mtune=native -march=native -mfpmath=both}.
Benchmarks are conducted using a workstation with two CPUs and a processor name of Intel(R) Xeon(R) Platinum 8276 CPU @ 2.20 GHz, which has 28 physical cores. Thus, there are 56 physical cores in total. The cache size of each CPU is 39,424 KB.
GPU benchmarks are performed with NVIDIA Tesla V100 SXM2 32GB, where the driver version is 440.100 and the CUDA version is 10.2. Since CUDA 10.2 is not compatible with GCC 9.3.0, binaries for GPU benchmarks are compiled with GCC 8.4.0.
The following benchmarks are performed with Python unless specified otherwise. To evaluate the execution times of Python functions, we use pytest~\cite{pytestx.y} with the default settings and use the minimum execution time for several runs. For C++ functions, we measure time with the \texttt{std::chrono} library. We repeat functions until the sum of times is 1 second or a function is executed $10^3$ times. Then we take use average time for the benchmark. 

\subsection{Performance of basic gates}
\begin{figure*}[t]
    \centering
    \begin{tabular}{c}
      \begin{minipage}[t]{14cm}
	\centering
    \includegraphics[width=\textwidth, clip]{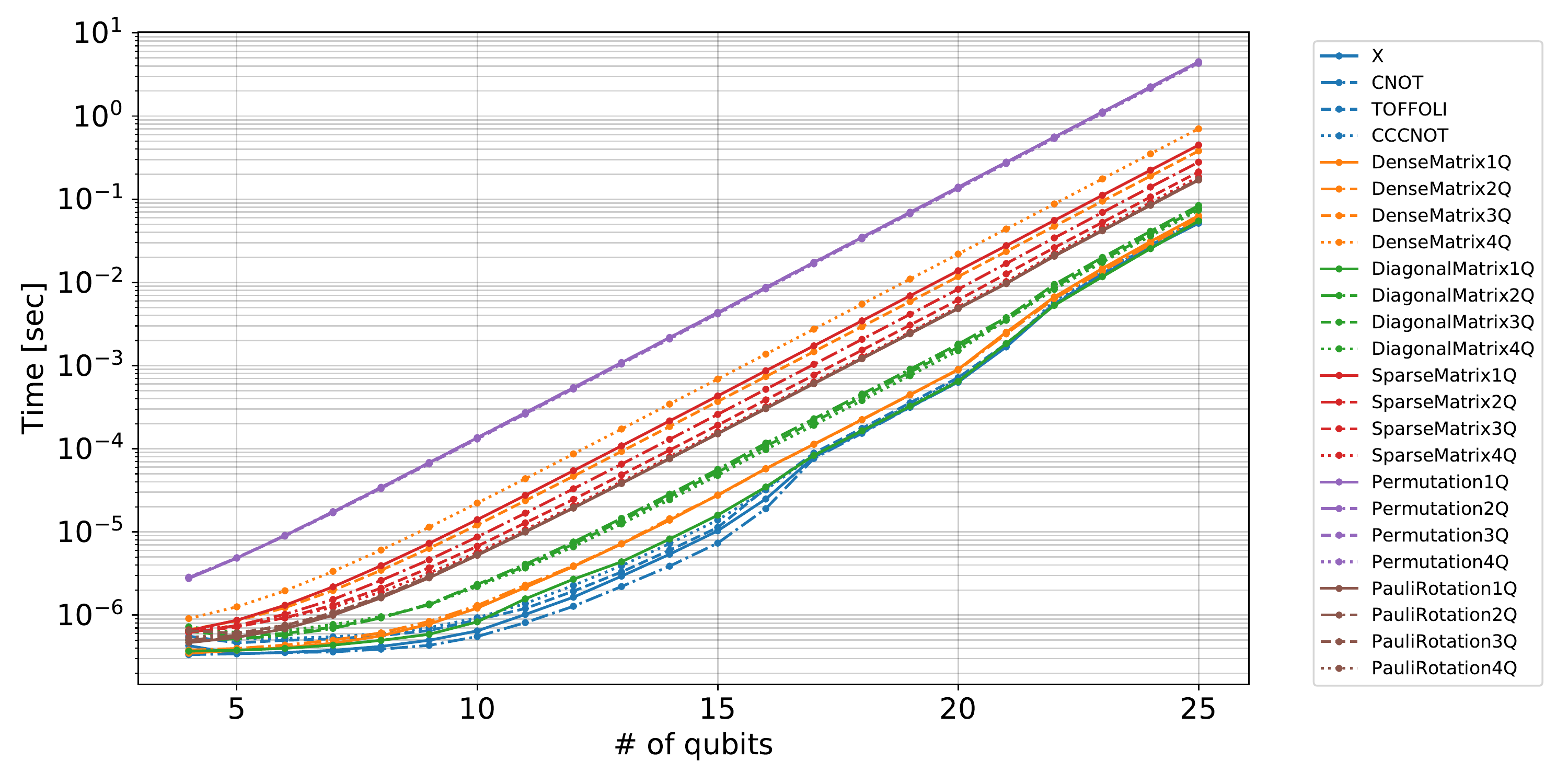}
	\caption{Times for applying basic gates are plotted as a function of the total number of qubits. The post-fix such as 1Q represents the number of target qubits. CCCNOT means a Pauli-$X$ gate controlled by three qubits.}
    \label{fig:compare_basic_gate}
      \end{minipage}
      \\
      \begin{minipage}[t]{14cm}
	\centering
    \includegraphics[width=\textwidth, clip]{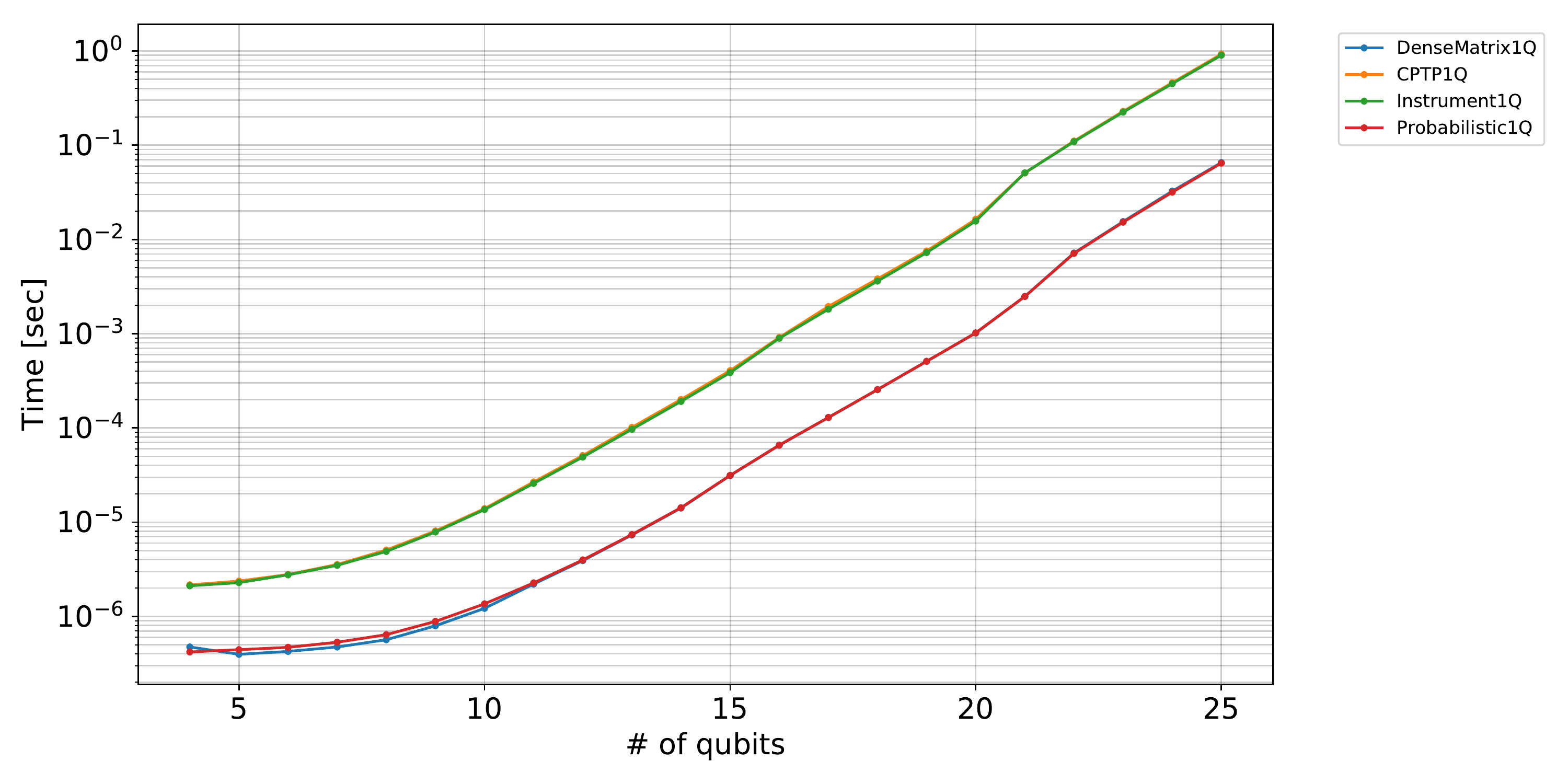}
	\caption{Times for applying general quantum maps are plotted as a function of the total number of qubits.}
    \label{fig:compare_general_gate}
      \end{minipage}
    \end{tabular}
\end{figure*}

First, we compare the times for applying basic gates via Python interfaces. Figure\,\ref{fig:compare_basic_gate} shows the time to apply gates as a function of the number of total qubits. 
The post-fix of legends such as Q1 represents the number of target qubits. Note that the time of the sparse matrix gate depends on the number of non-zero elements. We choose a matrix such that the top leftmost element is unity and the others are zero. 
All computational times grow exponentially with the number of qubits. When the number of qubits is small, the times for applying each quantum gate converge to a certain value, which is because of the overhead for calling C++ functions from Python. This overhead is evaluated in Sec.\,\ref{subsec:python_overhead}. 

The times for dense matrix gates also grow according to the number of target qubits $m$. Note that the times for $m=1,2$ are much faster than those for $m \geq 3$. This is because specific optimization is performed for dense matrix gates with $m=1,2$. They show almost the same values since the bottleneck in their simulation times is not due to arithmetic-operation costs but memory-operation costs, which are independent of the number of target qubits.

As expected, the times for diagonal matrix gates, Pauli rotation gates, and Permutation gates are independent of the number of target qubits $m$. Note that the time for diagonal matrix gate with $m=1$ is much faster than the other $m$ since specific optimization is performed for diagonal matrix gates with $m=1$. Thus, diagonal matrix gates should be used instead of dense matrix gates when the number of target qubits is large. Note that the times for permutation gates are much slower because a given reversible Boolean function is a Python function and its execution time has a large overhead compared with a function written in C++ language. Using the Qulacs as a C++ library should enhance the performance of permutation gates.

For sparse gates, the simulation time decreases according to the number of gates, as expected. 
Although the computational costs of controlled gates (i.e., Pauli-$X$, CNOT, TOFFOLI, and CCCNOT where CCCNOT is a bit-flip gate controlled by three qubits) should decrease as the number of control qubits increases, they do not show a clear scaling. This is because Pauli-$X$ and CNOT gates have SIMD optimized functions and because the bottleneck of their computing times is due to the memory-operation cost rather than the arithmetic-operation cost. Note that there are jumps at 17 and 22 qubits in the plots of controlled gates. This is because the memory size of the state vector exceeds the cache size, which causes discontinuous changes of the bandwidth for memory operations.

\subsection{Performance of general quantum maps}
Next, we compare the simulation times for dense matrix gates with general single-qubit quantum maps: CPTP-map, instrument, and probabilistic gates. We choose a quantum map such that one of two dense matrix gates is chosen with a probability of 0.5 as a benchmark target of a CPTP-map and probabilistic gates. We also choose the $Z$-basis measurement as that of an instrument. Figure\,\ref{fig:compare_general_gate} shows the benchmark results.
The probabilistic map has a similar performance to the dense matrix gate since it only requires overhead to randomly draw a gate whose probability is known in advance. On the other hand, CPTP-map and instrument are about 10 times slower. This is because they must not only calculate the probabilities of the Kraus operators but also allocate and release a temporal buffer to compute them. 

\subsection{Overhead due to a function call from Python}
\label{subsec:python_overhead}
\begin{figure*}[t]
    \centering
    \begin{tabular}{c}
      \begin{minipage}[t]{14cm}
	\centering
	\includegraphics[width=\textwidth, clip]{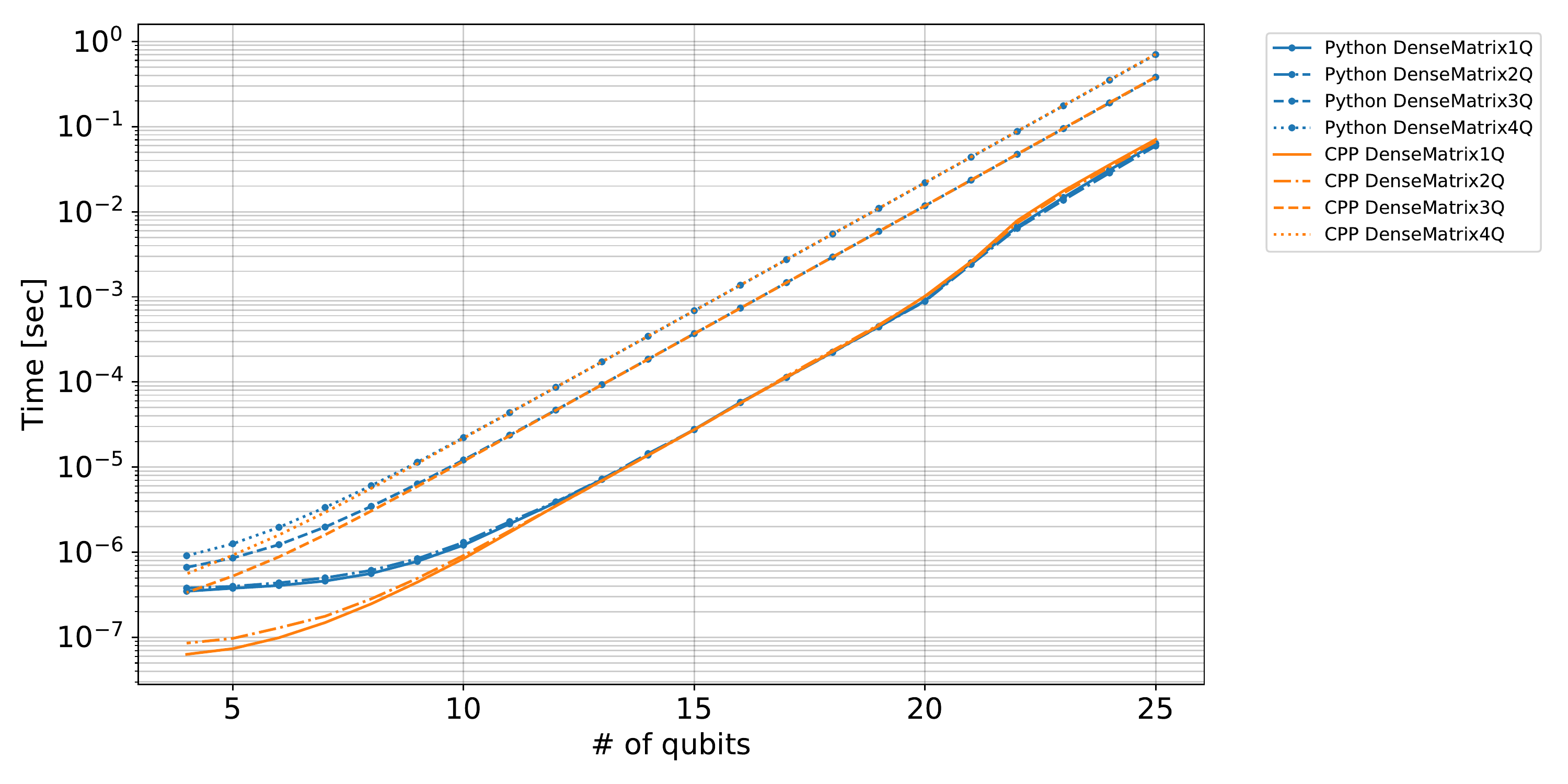}
	\caption{Times of function calls for applying dense matrix gates via python and C++ are compared.}
	\label{fig:compare_python}
      \end{minipage}
      \\
      \begin{minipage}[t]{14cm}
	\centering
	\includegraphics[width=\textwidth, clip]{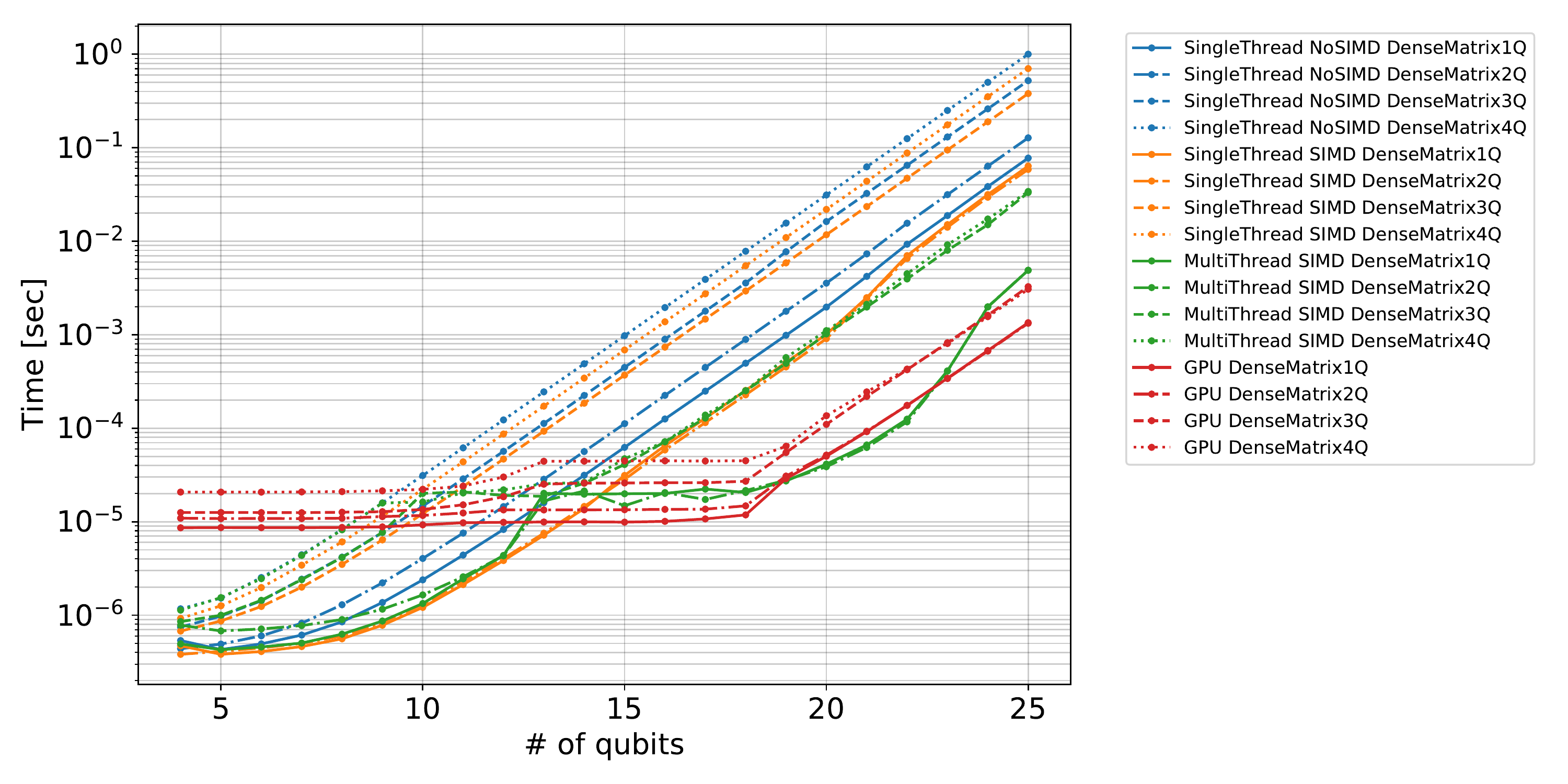}
	\caption{Times for applying dense matrix gates with several optimization settings are plotted as a function of the total number of qubits.}
	\label{fig:compare_gate_opt}
      \end{minipage}
    \end{tabular}
\end{figure*}

While Qulacs is written in C++ language, its functions can be called from Python. However, there is an overhead for calling a C++ function from Python. In the case of small quantum circuits, this overhead is not negligible. To evaluate this overhead, we compare the times for Qulacs with the Python interfaces to those for Qulacs without the Python interfaces. Figure\,\ref{fig:compare_python} shows the results. 
Although we made efforts to minimize the overhead due to Python interfaces, Qulacs still consumes about 0.3~${\rm \mu s}$ to call C++ functions from Python. This overhead is not negligible when the number of qubits is below 10. Thus, using Qulacs as a C++ library should increase the speed up to about seven times when the number of qubits is small.

\subsection{Speed-up by SIMD, OpenMP, and GPU}
We then evaluate performance improvement using the SIMD optimization, multi-threading with OpenMP, and GPU acceleration. We test the following settings: single-thread without SIMD, single-thread with SIMD, multi-thread with SIMD, and computing on a single GPU. 
Figure\,\ref{fig:compare_gate_opt} shows the times for applying dense matrix gates with several numbers of target qubits. 
The SIMD variants are always faster than the execution times without SIMD optimization. The OpenMP variant shows better performance when the number of qubits is greater than about 14 for $m=1,2$ and about 10 for $m=3,4$. Note that since multi-threading increases the overhead, and this overhead is not negligible in the case of a small number of qubits, Qulacs automatically uses the function without OpenMP in such cases. This is because the times for the OpenMP variant changes at a certain point. The GPU variants significantly improve the computing time when the number of qubits is large, but it requires about 10~${\rm \mu s}$ overhead. 

\subsection{Parallelization efficiency}
\begin{figure*}[t]
    \centering
    \begin{tabular}{c}
      \begin{minipage}[t]{14cm}
		\centering
	\includegraphics[width=\textwidth, clip]{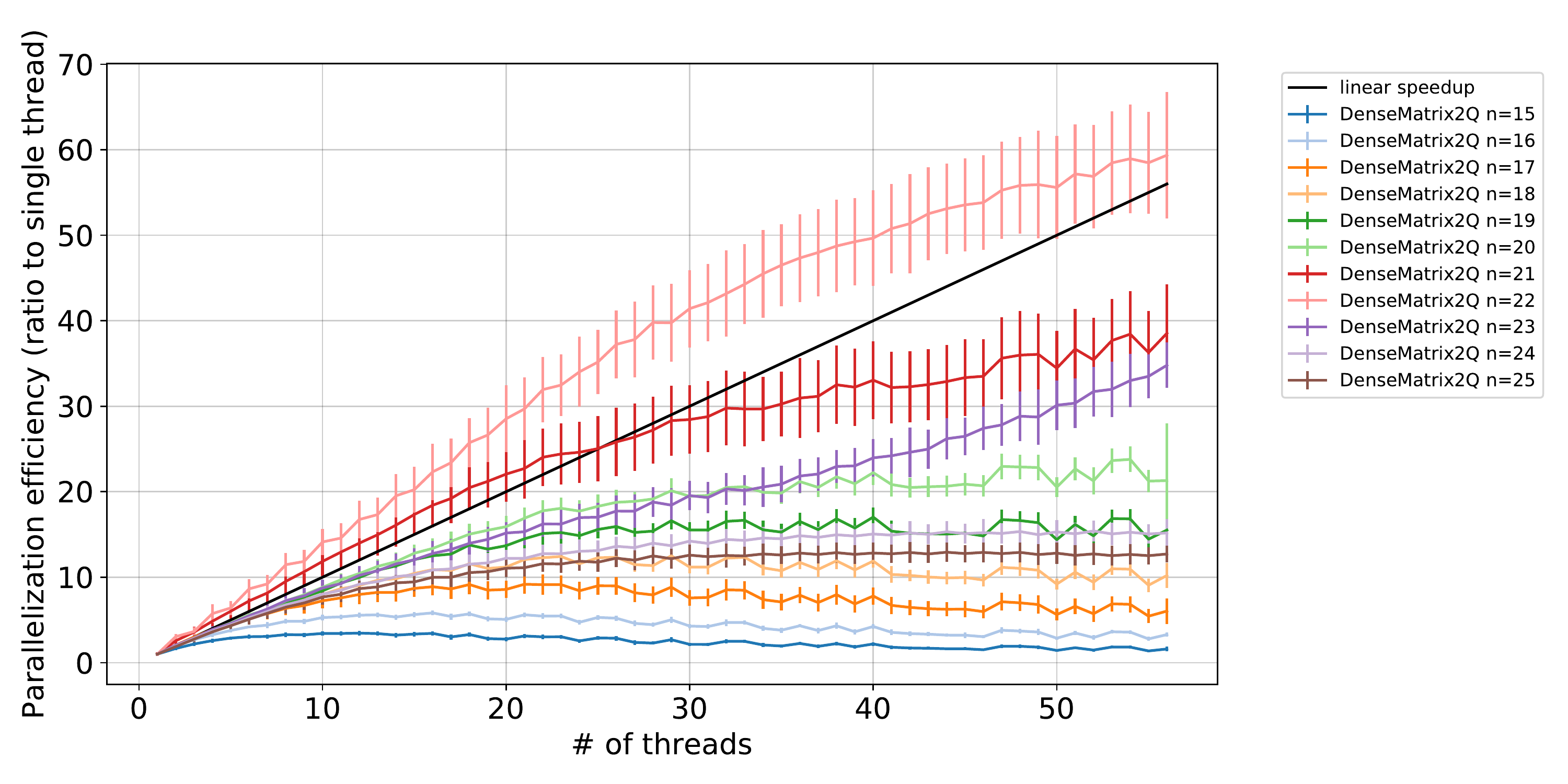}
	\caption{Times for applying dense matrix gates with several numbers of gates are plotted as a function of the total number of threads.}
	\label{fig:parallel_efficiency_auto}
      \end{minipage}
      \\
      \begin{minipage}[t]{14cm}
	\centering
	\includegraphics[width=\textwidth, clip]{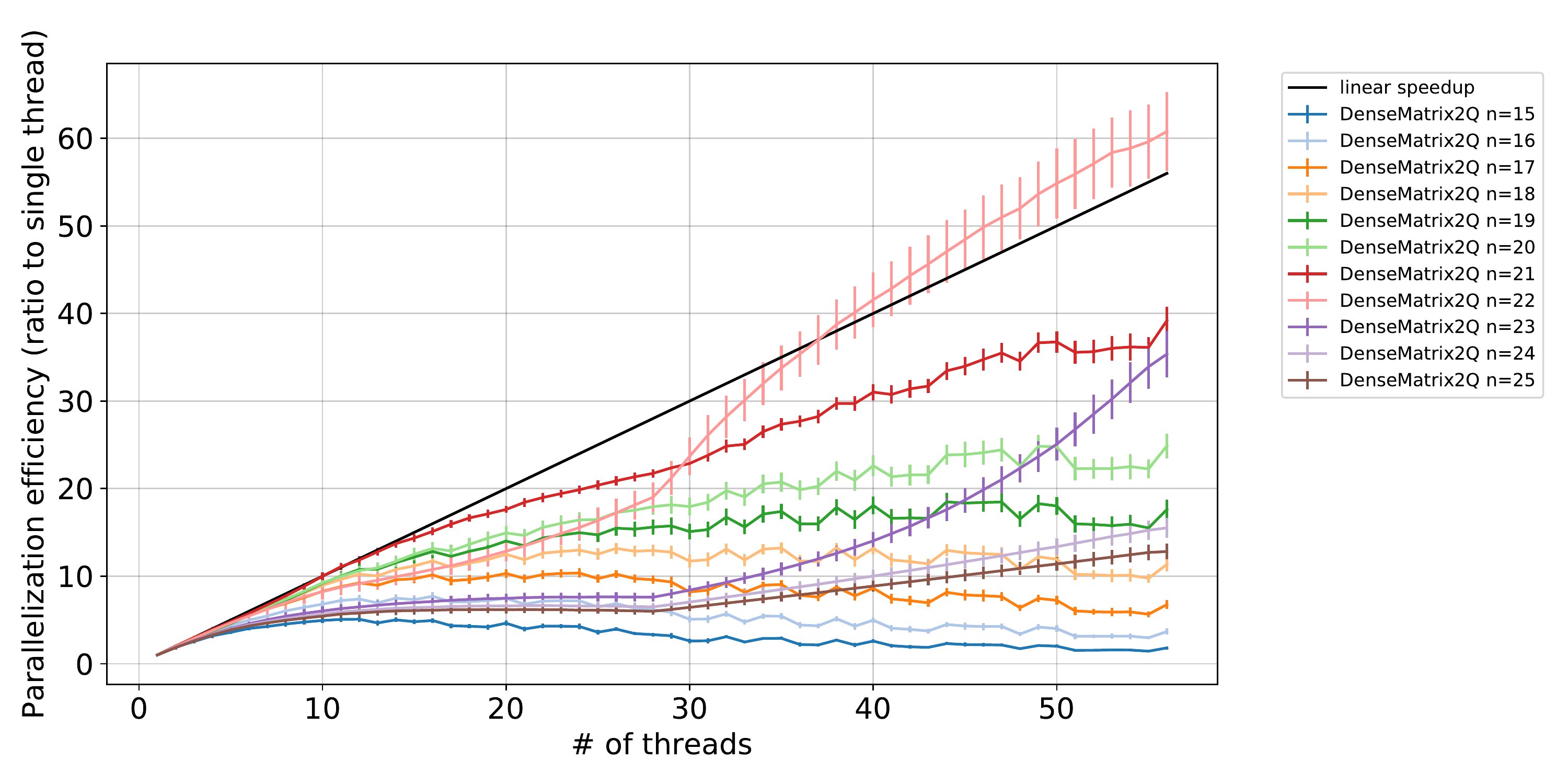}
	\caption{Times for applying dense matrix gates with several numbers of gates are plotted as a function of the total number of threads, where the setting of thread affinity is changed.}
	\label{fig:parallel_efficiency_close}
      \end{minipage}
    \end{tabular}
\end{figure*}

We discuss the parallelization efficiency of multi-threading with OpenMP. Figure\,\ref{fig:parallel_efficiency_auto} shows the execution time relative to the execution time of single-thread run with its standard error. We denote this ratio as parallelization efficiency, which becomes equal to the number of threads in the case of the linear speed-up. This ideal line is plotted as black lines in the figure. Since the overhead of parallelization becomes dominant when $n$ is small, Qulacs automatically disables the multi-threading when $n$ is smaller than about 13. Thus, we performed benchmark from $n=15$ to $n=25$. While we show the case of two-qubit gates, its behavior is almost the same as the case of single-qubit gates. Note that this evaluation is performed on the two CPUs workstation, each of which has 28 physical core, resulting in 56 physical cores in total. We vary the number of available threads from 1 to 56 with \texttt{OMP\_NUM\_THREADS}.

When the number of qubits is $n=15$, the parallelization efficiency saturates around thread number $t=10$, and then decreases as the number of threads increases due to the overhead of parallelization.
As the number of qubits increases, the granularity of the arithmetic and memory operations becomes coarser, and the parallelization efficiency improves gradually. Then, the performance trend drastically changes at $n=22$, where the performance improves beyond the linear speed-up. This can be explained with the size of cache and state vector; Our benchmark machine has two CPUs and each CPU has 40~MB L3 cache, and the size of the state vector is 67~MB at $n=22$, respectively. Thus, this is an exceptional situation where the whole state vector can reside in the L3 cache when we use two or more threads, assuming thread $N$ goes to CPU $N\%2$ (\texttt{OMP\_PLACES=socket}). Thus, in this situation, the effective bandwidth is super-linearly increased compared to the single-thread case. While the amount of speed-up depends on the index of quantum gates, we always observed super-linear speed-up at $n=22$ due to this reason.
To validate this explanation, we also performed the evaluation with \texttt{OMP\_PLACES=cores; OMP\_PROC\_BIND=close}. These environment variables force CPUs to use a single CPU until the number of threads is below the number of physical cores per CPU, and use two CPUs the number of threads exceeds it. Figure\,\ref{fig:parallel_efficiency_close} shows the performance with these options. We see that there is no improvement beyond the linear speed-up with a single CPU, and the performance drastically improves from $t>28$, which allows using the two CPUs. This behavior agrees with our explanation.

When the number of qubits becomes larger than $n=22$, the parallelization efficiency quickly reduces again. This is because the whole state vector cannot be stored in the cache in this region, and communication between the main RAM and CPUs is required. This reduces the effective bandwidth and the memory operation costs become dominant in the execution time. Since the multi-threading improves mainly the arithmetic operation costs, the advantage of parallelization becomes small.

\subsection{Circuit optimization}
\begin{figure*}[t]
    \centering
    \begin{tabular}{c}
      \begin{minipage}[t]{14cm}
	\centering
	\includegraphics[width=\textwidth, clip]{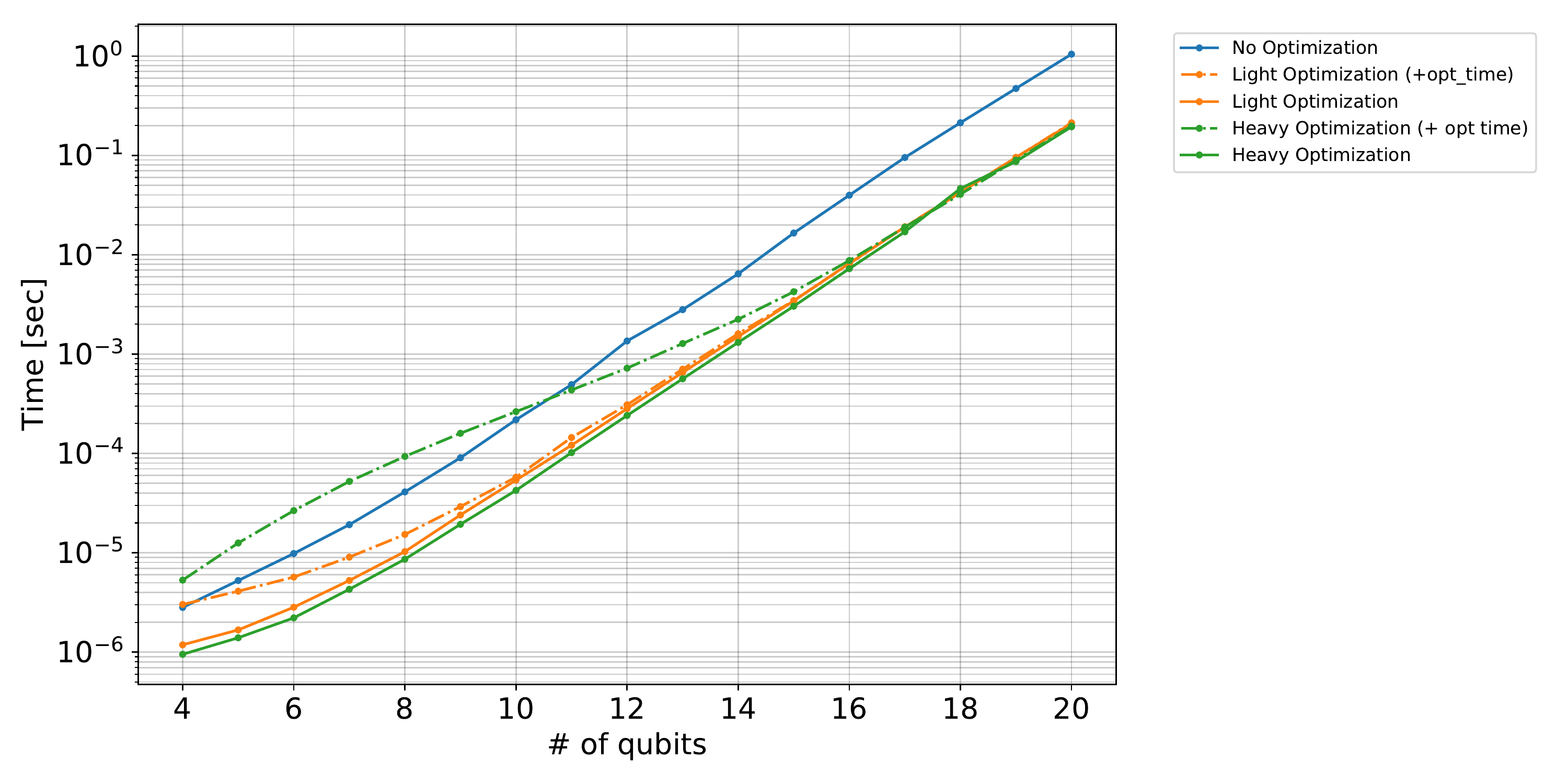}
	\caption{Times for simulating random quantum circuits are plotted with several strategies of quantum circuit optimization. "+ opt time" in legend represents that its plot includes a time for circuit optimization.}
	\label{fig:compare_circuit_opt}
      \end{minipage}
      \\
      \begin{minipage}[t]{14cm}
	\centering
	\includegraphics[width=\textwidth, clip]{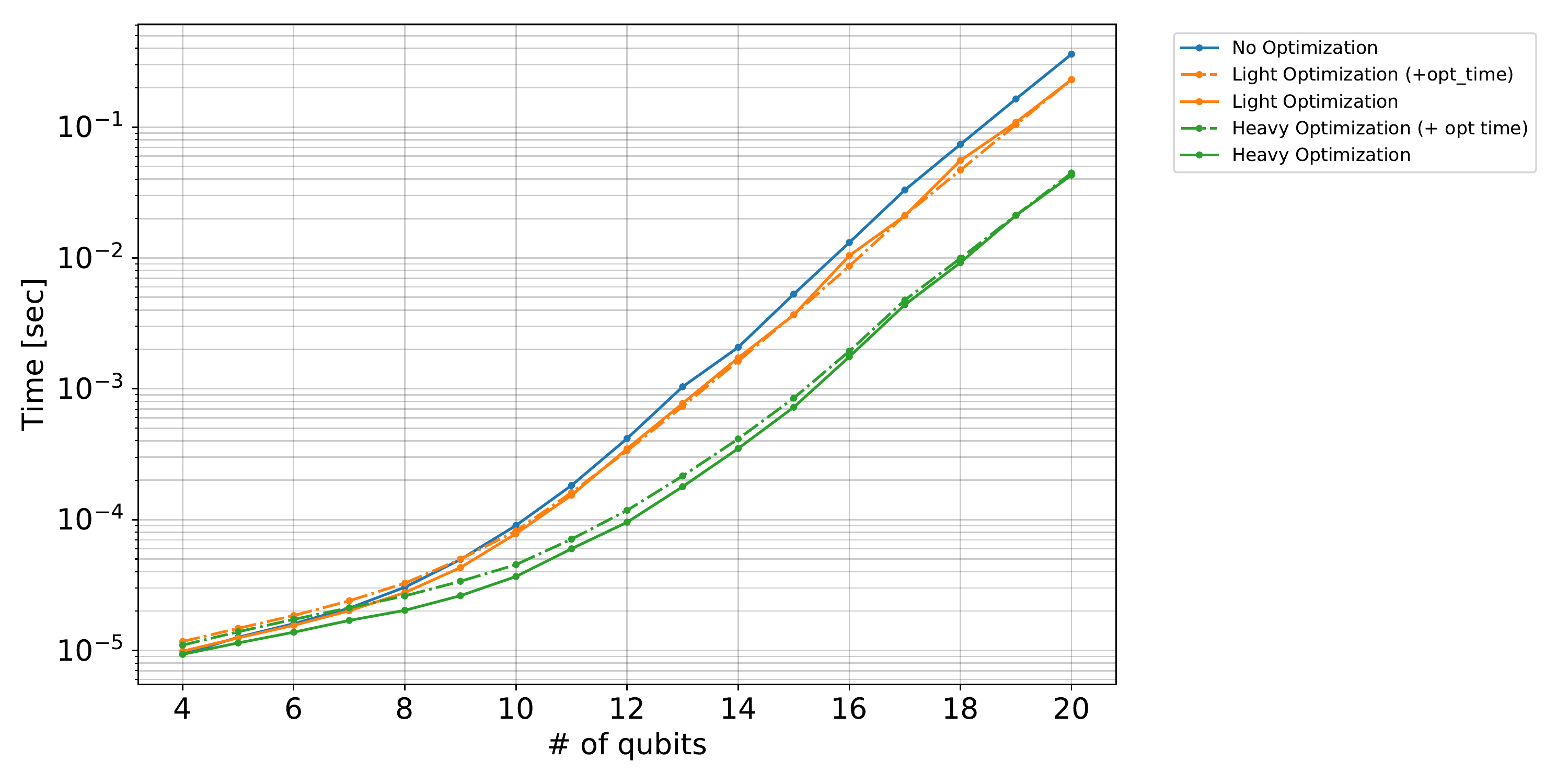}
	\caption{Times for simulating random quantum circuits dominated by commutative gates are plotted with several strategies of quantum circuit optimization.}
	\label{fig:compare_iqp_circuit_opt}
      \end{minipage}
    \end{tabular}
\end{figure*}

We evaluate the performance of two circuit optimization strategies. Here, we choose the following random quantum circuits for the benchmark. An $n$-qubit random quantum circuit consists of $n$ layers. In each layer, three random rotations (\texttt{RZ}, \texttt{RX}, \texttt{RZ}) act on each qubit, and controlled-$Z$ gates are applied to each neighboring qubit. Whether the starting index of the controlled-$Z$ gates is even or odd depends on the index of the layer. Finally, three random rotations act on each qubit. Listing.\,\ref{alg_randomqc} shows the source code to generate random circuits. 
\begin{lstlisting}[language=Python, caption=An Python program that generates random quantum circuits for our benchmark., label=alg_randomqc]
import numpy as np
from qulacs import QuantumCircuit
from qulacs.gate import RX, RZ, CZ
def generate_random_circuit(nqubits: int, depth: int) -> QuantumCircuit:
  qc = QuantumCircuit(nqubits)
  for layer_count in range(depth+1):
    for index in range(nqubits):
      angle1 = np.random.rand()*np.pi*2
      angle2 = np.random.rand()*np.pi*2
      angle3 = np.random.rand()*np.pi*2
      qc.add_gate(RZ(index, angle1))
      qc.add_gate(RX(index, angle2))
      qc.add_gate(RZ(index, angle3))
    if layer_count==depth:
      break
    for index in range(layer_conut%2, nqubits-1, 2):
      qc.add_gate(CZ(index, index+1))
  return qc
\end{lstlisting}
We simulate these circuits by enabling SIMD optimizations and disabling OpenMP. We choose a block size of two for the heavy optimizations. Figure\,\ref{fig:compare_circuit_opt} shows the simulation times for these circuits, where the solid and dashed lines exclude and include the circuit optimization time, respectively.
When the circuit optimization time is ignored, the performance of the heavy optimization is slightly better than that of the light optimization. However, when the optimization time is included in the computing time, there is no advantage in the heavy optimization. Thus, the light optimization (or no optimization) is more suitable in cases when a merged circuit is used only a few times.

Since the heavy optimization looks for pairs of quantum gates that can be merged considering commutation relations of gates, it is effective when there are many commutative gates in quantum circuits. When the three rotations (\texttt{RZ}, \texttt{RX}, \texttt{RZ}) are replaced with a single rotation \texttt{RZ}, all gates in quantum circuits become commutative in the $Z$-basis. In such a case, quantum circuits can be compressed to a constant depth by considering commutation relations. Figure\,\ref{fig:compare_iqp_circuit_opt} shows the simulation times for these circuits. 
As expected, the heavy optimization is effective in this case even if the optimization time is taken into account.

\section{Comparison with existing simulators}
\label{sec:comparison}
In this section, we compare the performance of Qulacs with that of existing libraries under the settings of single-thread, multi-thread, and with GPU acceleration.
We create a benchmark framework based on Ref.\,\cite{yaobenchmark}, of which the benchmark codes are reviewed by contributors of each library. This repository chooses the following random quantum circuits for the benchmark: Suppose we generate $n$-qubit random circuits. A layer with random \texttt{RZ, RX, RZ} rotations on each qubit is named a rotation layer. A layer with CNOT gates acting on the $i$-th qubit as a target qubit and $(i+1 \mod n)$-th qubit as a control qubit for ($0 \le i < n$) is named a CNOT layer. In a random circuit, a rotation layer and a CNOT layer are alternately repeated ten times, and a rotation layer follows it. Note that the first \texttt{RZ} rotations for all the qubits in the first rotation layer and the last \texttt{RZ} rotations in the last rotation layer are eliminated since they are meaningless if a quantum state is prepared in $\ket{0}^{\otimes n}$ and measured in the Pauli-$Z$ basis. See Refs.\,\cite{yaobenchmark, qulacsbench} for source codes for circuit generation.
The benchmarks with CPU are performed with a workstation with two CPUs and a processor name of Intel(R) Xeon(R) CPU E5-2687W v4 @ 3.00GHz on CentOS 7.2.1511. This workstation has 24 physical cores in total. The benchmarks with GPU are performed with two CPUs and a processor name of Intel(R) Xeon(R) Silver 4108 CPU @ 1.80 GHz and with Tesla V100 PCIe 32GB on CentOS 7.7.1908. The benchmarks are performed with double precision, i.e., each complex number is represented with 128 bits. Qulacs is compiled with the same options as those used in Sec.\,\ref{sec:performance}.
Versions of quantum circuit simulators and related libraries for CPU and GPU benchmarks are listed in Table.\,\ref{Tab:library_cpu} and Table.\,\ref{Tab:library_gpu}, respectively. 
\begin{table}[t]
\centering
\begin{tabular}{l|l}
Library & Version \\ 
\hline 
\hline
GCC & 9.2.0 \\
Python & 3.7.9 \\
Julia & 1.5.2 \\
NumPy & 1.19.2 \\
MKL & 2020.2 \\
TensoFlow & 2.3.1 \\
\hline
Intel-QS~\cite{guerreschi2020intel,smelyanskiy2016qhipster} & see main text \\
ProjectQ~\cite{steiger2018projectq} & 0.5.1 \\
PyQuEST-cffi~\cite{jones2019quest}  & 3.2.3.1 \\
Qibo~\cite{efthymiou2020qibo} & 0.1.2 \\
Qiskit~\cite{Qiskit} & 0.23.1 \\ 
Qiskit Aer~\cite{Qiskit} & 0.7.1 \\ 
Qiskit Terra~\cite{Qiskit} & 0.16.1 \\
Qulacs & 0.2.0 \\ 
qxelarator~\cite{khammassi2017qx} & 0.3.0 \\
Yao~\cite{Luo2020yaojlextensible} & 0.6.3 \\ 
\end{tabular}
\caption{A list of libraries and versions for CPU benchmark}
\label{Tab:library_cpu}
\end{table}
\begin{table}[t]
\centering
\begin{tabular}{l|l}
Library & Version \\ 
\hline 
\hline
GCC & 7.3.0 \\
Python & 3.7.9 \\
Julia & 1.5.2 \\
NumPy & 1.19.2 \\
MKL & 2020.2 \\
NVIDIA driver & 440.33.01 \\
CUDA & 10.2 \\
\hline
Qiskit~\cite{Qiskit} & 0.23.1 \\ 
Qiskit Aer~\cite{Qiskit} & 0.7.1 \\ 
Qiskit Aer GPU~\cite{Qiskit} & 0.7.1 \\ 
Qiskit Terra~\cite{Qiskit} & 0.16.1 \\
Qulacs & 0.2.0 \\ 
Yao~\cite{Luo2020yaojlextensible} & 0.6.3 \\ 
\end{tabular}
\caption{A list of libraries and versions for GPU benchmark}
\label{Tab:library_gpu}
\end{table}
All the simulators are installed with the latest stable versions as of November 2020. In benchmarks, a time for simulating a circuit to obtain the final state vector is evaluated, and a time for creating quantum circuits is not included. If an evaluated simulator offers the option of enabling circuit optimizations, we plot times both with and without quantum circuit optimization to discuss the advantage of acceleration by circuit optimization. Our benchmark codes can be found at Ref.\,\cite{qulacsbench}. 

While we carefully make the comparison fair, it is non-trivial to compare all the libraries in the exactly same conditions since they are developed with different purposes and designs. To clarify benchmark settings, here we describe several notes on how we install or evaluate simulators.
Qiskit~\cite{Qiskit} is a large software development kit, and Qiskit Aer is a component of Qiskit that implements fast simulation of quantum circuits. We used a backend named \texttt{StatevectorSimulator} for the benchmark since it is expected to be the fastest for simulating typical random circuits among the implemented simulators. In the default setting, we need to compile quantum circuits with \texttt{qiskit.compiler.transpile} and \texttt{qiskit.compiler.assemble} to create a job to call a core function of Qiskit Aer. A core function of Qiskit Aer is executed by submitting the compiled job to a thread pool. However, the overheads for these processes are sometimes longer than the time for simulation itself when the number of qubits is small. Thus, we eliminate the times for these processes from the benchmark and directly evaluate an execution time of a core function. Qiskit Aer performs circuit optimization similar to our technique, which is called gate fusion in Qiskit Aer. In the following discussion for Qiskit, we plot execution times with and without gate fusion, which can be switched via flag \texttt{enable\_fusion}. We note that execution times of \texttt{qiskit.execute} are longer than plotted ones.
Intel-QS~\cite{guerreschi2020intel}, which is also known as qHiPSTER, is a C++ library that has two official repositories in GitHub. In the benchmark, actively maintained one~\cite{intelqs} is used. In this repository, Intel-QS does not have stable release, so we installed the latest master branch of which the latest commit hash is \texttt{b625e1fb09c5aa3c146cb7129a2a29cdb6ff186a}. Intel-QS is compiled with GCC and with SIMD optimization, i.e., compiled with the options \texttt{CXX=g++} and \texttt{IqsNative=ON}. While Intel-QS is a C++ library, we use Intel-QS via python interface for a fair comparison.
QuEST~\cite{jones2019quest} is a C++ library of which the python interfaces are provided by another project named PyQuEST-cffi. While QuEST itself provides acceleration by parallelization and GPU, we skip the benchmark of QuEST with multi-thread and GPU acceleration since we cannot enable them via the python interfaces. 
ProjectQ~\cite{steiger2018projectq} raises errors to users when there is no measurement in circuits, which may cause an  additional overhead in benchmarks. While we expect this overhead is constant, we note that an execution time of ProjectQ may be faster than plotted when quantum circuits have measurements. 
Qibo~\cite{efthymiou2020qibo} is a library that supports GPU acceleration using TensorFlow, and we expect its performance depends on that of TensorFlow. However, since the latest TensorFlow is not compatible with CUDA 10.2, we skip the GPU benchmark of Qibo. In addition, since we installed TensorFlow with \texttt{pip} commands, TensorFlow does not support AVX2 extension. Thus, the performance of Qibo in the CPU benchmark would be a few times faster than plotted ones by installing TensorFlow from the source.
QX Simulator~\cite{khammassi2017qx} is used with its python interface named qxelerator, which accepts a file with QASM-format~\cite{cross2017open} strings as an input, we generate a benchmark circuit, convert it to a QASM string, save it as a file, load it with qxelerator, and evaluated a time for simulation, i.e., a time for executing \texttt{qxelerator.QX.execute} function.
Since QCGPU and qsim only support simulation with single precision, we did not perform the benchmarks for them. 
For all the benchmark libraries, their execution times may vary depending on the structure of quantum circuits for benchmarks, library versions, compilers, the status of the CPU and GPU, simulation environment, etc. In particular, while execution times for single-thread simulation are stable, those for multi-thread and GPU fluctuate by a few percent each time we run a benchmark script. Thus, it should be noted that a few percent difference in multi-thread and GPU benchmarks are meaningless. It should be also noted that several libraries among the above such as Intel-QS, QuEST, QX Simulator, and ProjectQ support quantum circuit simulation with distributed computing and are not necessarily optimized for the single-node performance.

\begin{figure*}[t]
    \centering
    \begin{tabular}{c}
      \begin{minipage}[t]{14cm}
		\centering
	\includegraphics[width=\textwidth, clip]{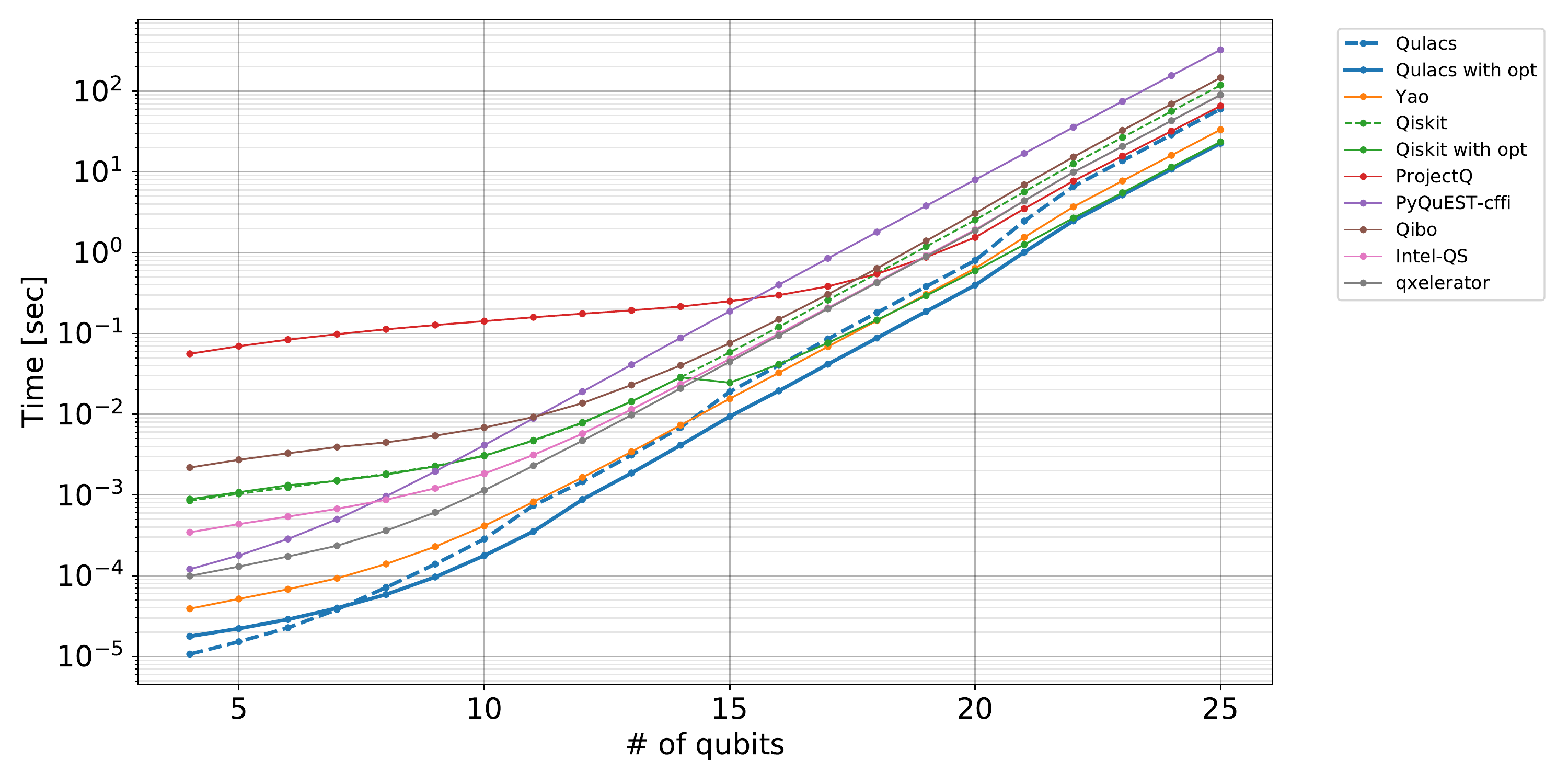}
	\caption{Times for simulating random quantum circuits with a single thread using several libraries.}
	\label{fig:compare_library}
      \end{minipage}
      \\
      \begin{minipage}[t]{14cm}
	\centering
	\includegraphics[width=\textwidth, clip]{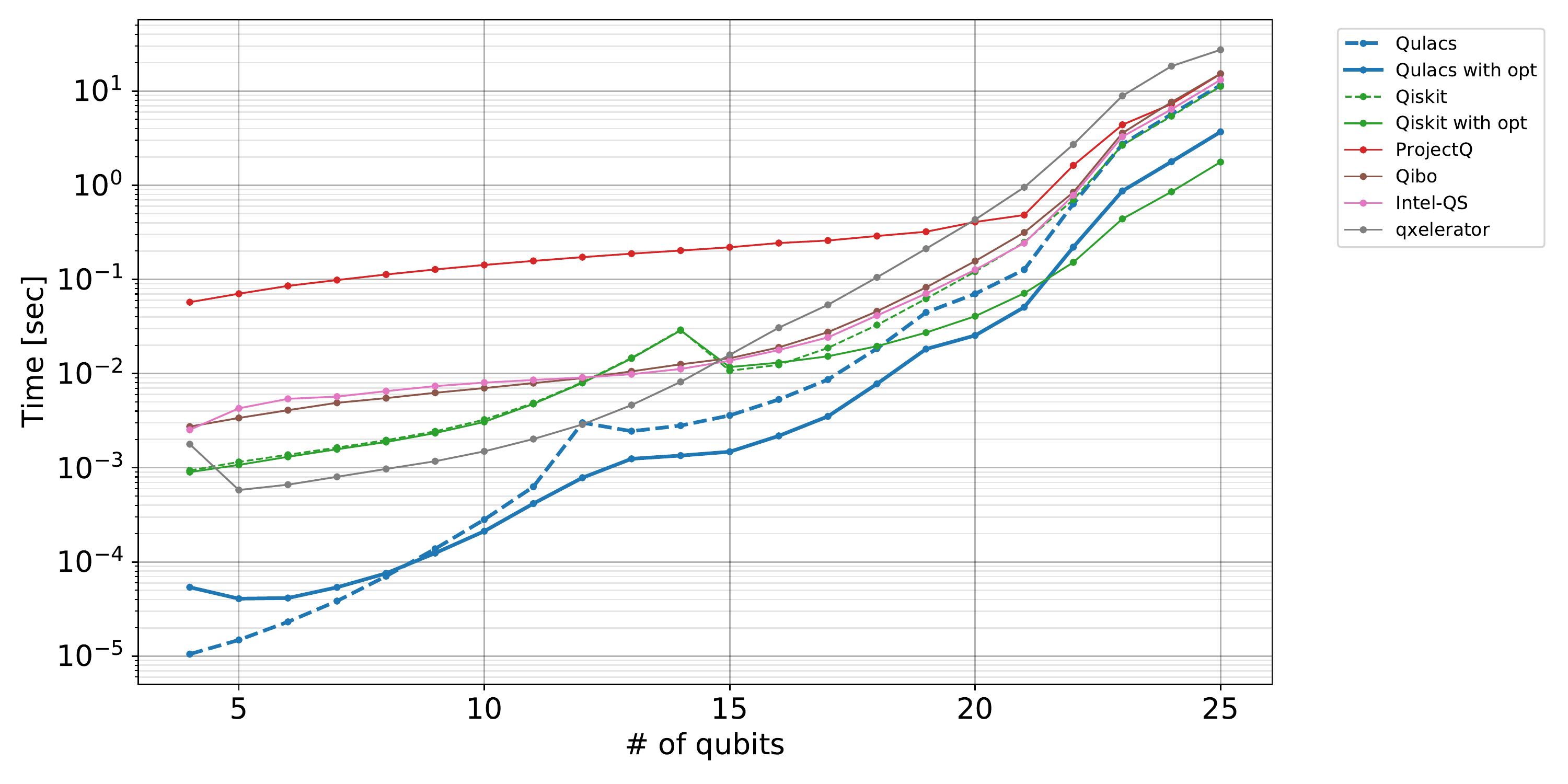}
	\caption{Times for simulating random quantum circuits with parallelization using several libraries.}
	\label{fig:compare_library_parallel}
      \end{minipage}
    \end{tabular}
\end{figure*}

First, we perform benchmarking with a single thread simulation with CPU. Figure\,\ref{fig:compare_library} shows the results.
For Qulacs and Qiskit, the performance of them with and without circuit optimization is plotted as a bold line and broken line, respectively. Qulacs without optimization is faster than that with optimization when the number of qubits is small. On the other hand, Qulacs with optimization becomes faster than Qulacs without optimization when the number of qubits is greater than 7 since the time for circuit optimization becomes negligible compared to the simulation time.
As far as we know, Qiskit with optimization and Yao utilize the structure of quantum circuits to improve the performance, and thus their performance overcomes that of Qulacs without circuit optimization.
Since Qiskit without optimization is slower than Qulacs without optimization but they are comparable with optimization, we guess circuit optimization by Qiskit is more time-consuming but near-optimal than that by Qulacs. 

Second, we perform benchmarking with multi-thread computing. Figure\,\ref{fig:compare_library_parallel} shows the results.
Due to a small overhead of Qulacs, its execution time is the fastest when the number of qubits is small. When the number of qubits becomes large, several libraries without circuit optimization achieve almost the same performance. When multi-threading is enabled, an execution time is determined by memory operations rather than arithmetic operations. Since the number of memory operations is almost independent of the detail of implementation, it is natural that times of several libraries converge to a certain value. 
When we compare the performance of libraries with circuit optimization, Qulacs with optimization shows better performance than that without optimization above 9 qubits. Since Qiskit is expected to perform near-optimal circuit optimization, Qiskit shows better performance than Qulacs when the number of qubits is larger than 21.
Note that there is a small bump around 14 qubits in the plot of Qiskit, which happens because Qiskit enables multi-thread when the number of qubits is more than 14 qubits in the default setting, thus the times of Qiskit around 14 qubits can be slightly faster by optimization of settings.

\begin{figure*}[t]
    \centering
    \begin{tabular}{c}
      \begin{minipage}[t]{14cm}
	\centering
	\includegraphics[width=\textwidth, clip]{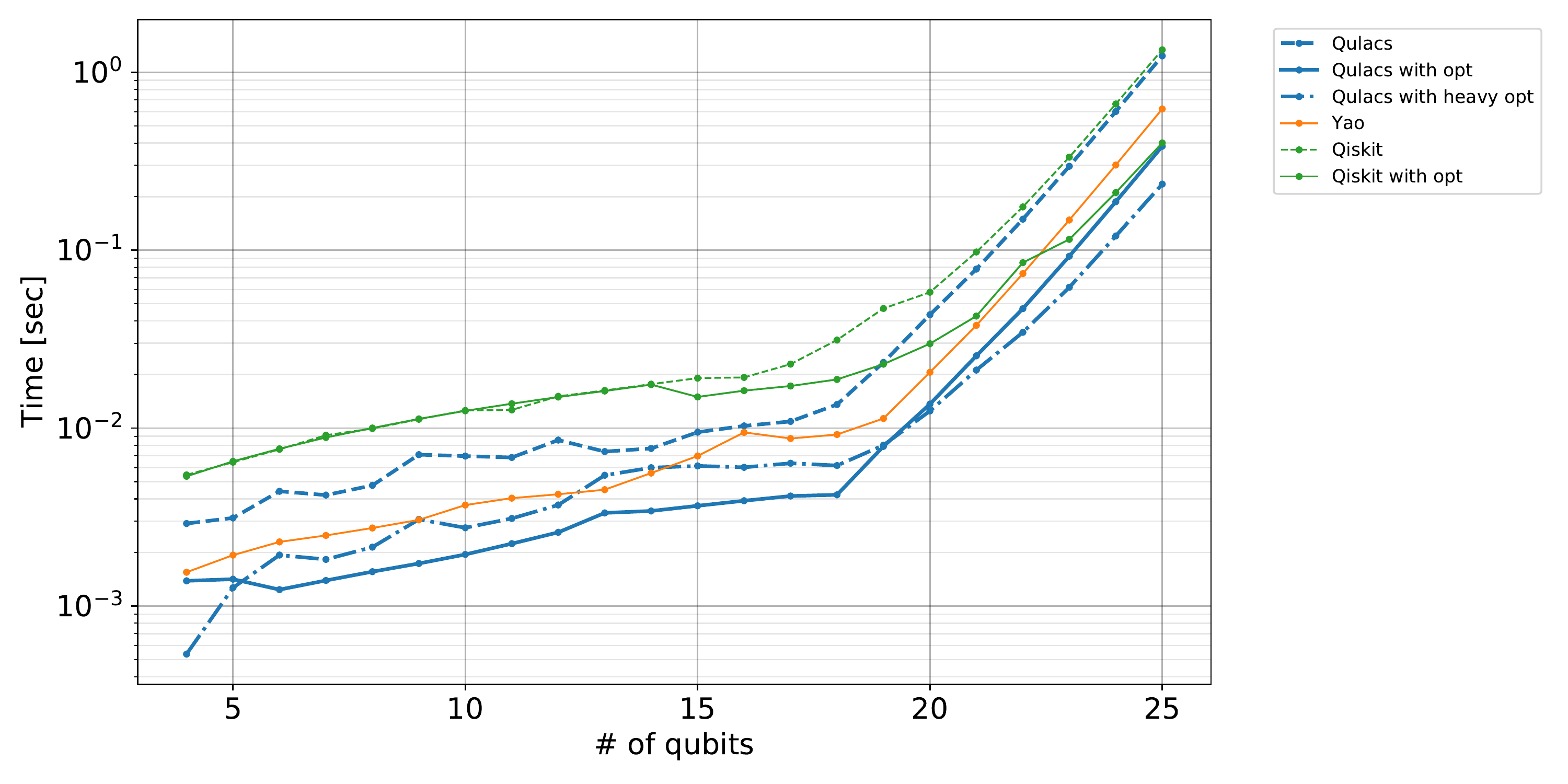}
	\caption{Times for simulating random quantum circuits with GPU acceleration using several libraries.}
	\label{fig:compare_library_gpu}
	\end{minipage}
    \end{tabular}
\end{figure*}

Finally, we perform the benchmark with GPU acceleration. Figure\,\ref{fig:compare_library_gpu} shows the results.
In the GPU benchmarking, Qulacs is also one of the fastest libraries. In the GPU simulation, the overhead due to circuit optimization becomes negligible since there is a larger overhead due to GPU function calls. Therefore, the performance of Qulacs with circuit optimization is always better than that without circuit optimization. Since times for circuit optimization is negligible, we can further improve the performance of Qulacs by utilizing the heavy optimization. As far as we tried, the heavy optimization with the block size of four is optimal. The performance with the heavy optimization is plotted in the figure with the legend "Qulacs with heavy opt". The performance of Qulacs with the heavy optimization overcomes that with the light optimization above 19 qubits. 
Note that the time for Qulacs with the heavy optimization at $n=4$ is significantly faster than the other number of qubits. This is because a circuit is merged to a single quantum gate by an optimizer and a GPU function is called at once.

In conclusion, Qulacs is one of the fastest simulators among existing libraries in several regions that are vital for researches. In particular, Qulacs shows significant speed-up when the number of qubits is small, which is essential for exploring the possibilities of quantum computing.

\section{Conclusion and Outlook}
\label{sec:conclusion}
Here, we introduced Qulacs, which is a fast and versatile simulator. First, we showed the basic concept and intended usages of Qulacs. Second, we explained the library structure and provided several examples. We optimized the update functions of Qulacs according to the properties of gate matrices. We then utilized additional optimization techniques such as SIMD optimization, multi-threading with OpenMP, GPU acceleration, and circuit optimization for numerical speed-up. We also showed concrete simulation times for several quantum gates and evaluated speed-up by optimization techniques. Finally, we compared the performance of Qulacs with that of the existing libraries. With the benchmarks, we showed our simulator has advantages in several scenarios. Although Qulacs focuses on supporting fundamental operations, we can use Qulacs to explore simulations with many layers using it as a backend of other libraries, for example, Cirq~\cite{quantum_ai_team_and_collaborators_2020_4062499} and OpenFermion~\cite{mcclean2020openfermion}. 

When quantum circuits are constructed only with efficiently simulatable quantum gates such as Clifford gates~\cite{gottesman1998heisenberg,aaronson2004improved} or matchgates~\cite{valiant2002quantum,terhal2002classical,knill2001fermionic}, these circuits are efficiently simulated via specialized algorithms. We plan to implement these algorithms and support faster simulations of quantum circuits in the future.

\section*{Acknowledgment}
Yasunari Suzuki, Yoshiaki Kawase, Yuria Hiraga, Yuya Masumura, Masahiro Nakadai, and Keisuke Fujii are the core contributors of Qulacs.
Tennin Yan, Yohei Ibe, Toru Kawakubo, Hirotsugu Yamashita, Hikari Yoshimura, Jiabao Chen, and Youyuan Zhang have made significant efforts to maintain manuals, repositories, and web sites.
Ken M. Nakanishi, Kosuke Mitarai, Yuya O. Nakagawa, Shiro Tamiya, Takahiro Yamamoto, Ryosuke Imai, and Akihiro Hayashi provided many essential comments and directions for improving the performance and usability of Qulacs. 

Yasunari Suzuki would like to thank Tyson Jones for the fruitful discussion on parallel and distributed computing, Xiu-Zhe Luo for the vital comments on vectorization, and Shinya Morino for the advice on GPU acceleration. We would also like to thank all the contributors and users of Qulacs for supporting this project.

This work is supported by PRESTO, JST, Grant No.\,JPMJPR1916; ERATO, JST, Grant No.\,JPMJER1601; MEXT Q-LEAP Grant No.\,JPMXS0120319794, JPMXS0118068682, and JPMXS0120319794.

\appendix
\section{C++ example codes}
\label{appendix:impl_cpp}
To show Qulacs can be used as a C++ library in almost the same way as Python, we show the example codes of Qulacs in the C++ language. Listing.\,\ref{alg_cpp} shows an example procedure to create, modify, update, and release quantum states using quantum gates. This program outputs a message shown in Listing.\,\ref{alg_cpp_out}. As you can see, the names and design of API are almost the same as the python library, and we can use Qulacs as a C++ library with small difference, e.g., complex matrices are supplied using Eigen instead of NumPy, users have a responsibility to release allocated state vector, and so on. For more detailed examples, see the online manual of Qulacs~\cite{Qulacs}.

\begin{lstlisting}[language=C++, caption=An example C++ program that initializes quantum states., label=alg_cpp]
#include <vector>
#include <complex>
#include <Eigen/Core>
#include <cppsim/state.hpp>
#include <cppsim/gate_matrix.hpp>
#include <cppsim/gate_factory.hpp>

int main(){
  unsigned int num_qubit = 2;

  // create state vector
  QuantumState* state = new QuantumState(num_qubit);
  state->set_computational_basis(2);
  QuantumState* sub_state = state->copy();

  std::vector<std::complex<double>> values = {0.5, 0.5, 0.5, -0.5};
  state->load(values);
  state->load(sub_state);
  state->set_Haar_random_state(42);

  Eigen::MatrixXcd gate_matrix(4,4);
  gate_matrix <<
          1, 0, 0, 0,
          0, 1, 0, 0,
          0, 0, 0, 1,
          0, 0, 1, 0;
  QuantumGateBase* dense_gate = gate::DenseMatrix({0, 1}, gate_matrix);
  dense_gate->update_quantum_state(state);

  QuantumGateBase* swap_gate = gate::SWAP(0,1);
  swap_gate->update_quantum_state(state);
  std::cout << dense_gate << std::endl;
  std::cout << swap_gate << std::endl;
  std::cout << state << std::endl;

  delete dense_gate;
  delete swap_gate;
  delete state;
  delete sub_state;
}
\end{lstlisting}

\begin{lstlisting}[caption=An output message of the program shown in Listing.\,\ref{alg_cpp}, label=alg_cpp_out]
 *** gate info ***
 * gate name : DenseMatrix
 * target    :
 0 : commute
 1 : commute
 * control   :
 * Pauli     : no
 * Clifford  : no
 * Gaussian  : no
 * Parametric: no
 * Diagonal  : no
 * Matrix
(1,0) (0,0) (0,0) (0,0)
(0,0) (1,0) (0,0) (0,0)
(0,0) (0,0) (0,0) (1,0)
(0,0) (0,0) (1,0) (0,0)

 *** gate info ***
 * gate name : SWAP
 * target    :
 0 : commute
 1 : commute
 * control   :
 * Pauli     : no
 * Clifford  : yes
 * Gaussian  : no
 * Parametric: no
 * Diagonal  : no

 *** Quantum State ***
 * Qubit Count : 2
 * Dimension   : 4
 * State vector :
   (0.343453,0.418285)
 (-0.287486,-0.663243)
  (0.309688,-0.266554)
(-0.0511105,-0.122348)
\end{lstlisting}

\bibliographystyle{unsrtnat}
\bibliography{bibQulacs}

\end{document}